\documentclass[pra,preprint, superscriptaddress,amsmath,amssymb,floatfix]{revtex4-2}

\usepackage{graphicx}
\usepackage{bm}
\usepackage[colorlinks=true,linkcolor=blue,urlcolor=blue,citecolor=blue,pdfstartview=FitH]{hyperref}
\usepackage{amsmath}
\usepackage{microtype}
\usepackage{siunitx}

\usepackage{xr-hyper}
\usepackage{hyperref}
\newcommand\norm[1]{\left\lVert#1\right\rVert}

\begin{document}



\title{Synchronization induces Bell violations in a model of walking droplets}


\author{Álvaro G. López}
\affiliation{Nonlinear Dynamics, Chaos and Complex Systems Group.\\Departamento de F\'isica, Universidad Rey Juan Carlos, Tulip\'an s/n, M\'ostoles, 28933, Madrid, Spain}\email{alvaro.lopez@urjc.es}
\author{Rahil N. Valani}
\affiliation{Rudolf Peierls Centre for Theoretical Physics, Parks Road,
University of Oxford, OX1 3PU, United Kingdom}\email{rahil.valani@physics.ox.ac.uk}
\author{Yuanmei Li}
\affiliation{National Demonstration Center For Experimental Physics Education, School of Physics, Nanjing University, 22 Hankou Road, Gulou District, Nanjing, Jiangsu 210093, China}
\author{John W. M. Bush}
\affiliation{Department of Mathematics, MIT, 77 Massachusetts Ave., Cambridge MA 02139, USA}\email{bush@math.mit.edu}



\date{\today}

\begin{abstract}
We consider a reduced Lorenz-like model that describes two walking droplets interacting through their mutual wave field, and investigate the emergence of strong bipartite correlations in this classical wave–particle system. The coupled nonlinear dynamics admit two invariant synchronization manifolds associated with correlated and anticorrelated states, within which the droplets display synchronized chaotic intermittency. Employing measurement protocols inspired by Bell experiments, we compute position correlations from {the long-time} dynamics and identify parameter regimes for which the CHSH-Bell parameter $S$ exceeds 2, corresponding to violations of Bell's Inequality. We further introduce a procedure for isolating the two subsystems, thereby ensuring the absence of wave-mediated signaling between them. Changing measurement settings following this isolation allows us to execute dynamic Bell tests in which violations persist. Our results demonstrate that nonlinear deterministic dynamics can produce {Bell violations through wave-mediated} synchronization mechanisms; {moreover, these violations may be rationalized on the grounds that the wave form is influenced by the measurement settings}. We thus provide a consistent dynamical framework for the {appearance} of classical entanglement in pilot-wave systems.
\end{abstract}

\maketitle

\section{Introduction}

Quantum mechanics describes the statistical behavior of microscopic particles. Hidden variable theories would complete this statistical description by providing a prescription for the dynamics of quantum particles. Impossibility proofs in quantum mechanics were developed with a view to informing the viability of hidden-variable theories \cite{Brukner2012Bell}. Such impossibility proofs have an inauspicious history \cite{BricmontBook}. In 1932, von Neumann presented an impossibility proof that was alleged to rule out all hidden-variable theories, with the inference that the statistical description of quantum systems was complete. While the flaw in von Neumann’s proof was pointed out shortly thereafter by Grete Hermann \cite{Hermann1935}, its deficiencies were not widely recognized for 20 years, until Bohm presented a counterexample: a theory of quantum dynamics that is consistent with quantum statistics \cite{Bohm1952a}. This counterexample attracted the attention of Bell \cite{Bell1964}, who wrote widely on the subject of quantum foundations \cite{Bell1987}, debunked the extant impossibility proofs, and concluded that `the only thing proven by impossibility proofs is the author’s lack of imagination’.  Despite this warning, Bell’s enduring legacy was an impossibility proof of his own.

Bell’s theorem is an inequality that bounds the degree of correlation that may arise between measurements of a dichotomic property (typically spin -- up or down) of two particles in a bipartite system. In its derivation, three fundamental assumptions are explicitly made. First, it is assumed that an external reality exists independent of human observation. Second, it is assumed that there is no superluminal signaling between the bipartite systems, that the universe is local. Third, {it is assumed that the hidden variables are independent of the measurement settings, an assumption referred to as} \emph{measurement independence}. The logical inference from the experimental violation of Bell’s inequality by Aspect and coworkers \cite{Aspect1981, Aspect1982a, Aspect1982b} is that one of the assumptions made in its derivation is false. It is widely believed that these violations preclude the possibility of local hidden variables; ergo, quantum dynamics is nonlocal \cite{Maudlin2014}. 

Quantum pilot-wave theories seek a quantum dynamics in which a particle moves in response to a guiding wave field. In Bohmian mechanics \cite{Bohm1952a}, this wave field is the wave function that characterizes the system's statistical behavior. According to de Broglie’s theory of the double-solution~\cite{deBroglie1987}, the particle has an intrinsic vibration that serves as the source of its guiding wave. The resulting pilot-wave dynamics was posited, but never proven to give rise to statistical behavior consistent with that described by the standard theory. It was referred to as the double solution because it entailed prescribing two distinct waves: the physical wave that guides the particle and the emergent statistical wave. The same physical picture arises in stochastic electrodynamics (SED), where the pilot wave is sought in the electromagnetic quantum vacuum \cite{Pena1996,Pena2015}. In Bohmian mechanics, the hidden variables are the particle's position and momentum \cite{Holland1993}. In de Broglie's pilot-wave theory and SED, these hidden variables must be augmented by those characterizing the form of the real electromagnetic pilot wave \cite{Lopez2020}, which one expects to depend on the local system geometry or measurement settings. Vervoort has therefore questioned the validity of the assumption of measurement independence for pilot-wave systems \cite{Vervoort2016a, Vervoort2018}.

In 2005, Couder discovered that a millimetric droplet may self-propel along the surface of a vibrating bath, guided by its own wave field ~\cite{Couder2005a}. Pilot-wave hydrodynamics is the field concerned with the dynamics and emergent statistics of this system \cite{Bush2015}. In many settings, the emergent statistics are similar to those that arise in quantum systems.  Examples include coherent diffraction patterns arising from walkers interacting with slits \cite{Couder2006, Pucci2018, ellegaard2020interaction} and standing waves \cite{primkulov_diffraction_2025, Tadrist-KD2026}, as well as robust wave-like statistics
emerging for droplets walking in circular or elliptical corrals~\cite{Harris2013, Saenz2018}. When external forces confine the walker's motion, the droplet tends to follow quantized orbits; when chaos ensues, the droplet intermittently switches between some finite number of orbits, evoking the notion of a superposition of states~\cite{Oza2014,Harris2014a,Perrard2014}. This pilot-wave hydrodynamic system represents a macroscopic realization of de Broglie’s double-solution theory \cite{deBroglie1987}, in that it is characterized by both a real guiding wave and an emergent statistical wave. Its success as a platform for analogizing quantum systems, and so for the field of hydrodynamic quantum analogs, would seem to lend support to de Broglie’s theory and its modern extension, stochastic electrodynamics~\cite{Pena2015}. A key feature of the walking-droplet system is that the bath serves as the system's memory \cite{Couder2012}; specifically, the history of the droplet and its environment is encoded in its wave field. Quantum features arise in the long-memory limit when the waves are most persistent. The field of hydrodynamic quantum analogs naturally raises the question: ``Might memory account for entanglement?" \cite{Bush2010}.

The interactions between walking droplets have been considered in several contexts ~\cite{GaleanoRios2018,Couchman2019,Valani2018, Valani2018_2, Papatryfonos2022,CorrNachbin2026SpontaneousOW}. Nachbin~\cite{correlationnachbin} examined particle-particle correlations that may arise in walking-droplet pairs, established via wave-mediated forces. He considered the motion of a pair of droplets confined to {one-dimensional} wells, free to move within their respective wells and communicated via a wave field spanning both wells. In addition to regimes marked by perfect periodic synchronization, he identified regions in which the particles exhibited asynchronous chaotic dynamics yet were statistically indistinguishable: their position-momentum phase-space probability distributions were identical. He also found that synchronization between the droplets could persist even after the two subsystems were isolated: the droplets remained in a dynamical state marked by the wave-mediated memory of their partner \cite{Nachbin2022}.

Bell violations have been reported in several classical systems, but have predominantly occurred in pure wave systems, either optical~\cite{Goldin,Qian} or acoustic~\cite{sound_Bell_2019}.
Papatryfonos \emph{et al}.~\cite{PhysRevFluids.9.084001} 
simulated the dynamics of a pair of droplets, each in a subsystem consisting of two wells between which the droplet could tunnel. These two subsystems were linked by a coupling cavity, which was forbidden to the droplets, that allowed waves to pass between them. A static Bell test was executed by identifying the drop location (inner or outer well) with the dichotomic property (spin up or down) and the depth of the tunneling barrier with the measurement setting (polarizer angle) in the optical Bell tests. The system was found to be ergodic, so the initial conditions were not determinant: the system could settle into periodic or chaotic states. Through a judicious choice of measurement settings, violations of Bell’s inequality were achieved. The emergent correlations and associated Bell violations were rooted in wave-mediated interactions between the two subsystems. The Bell violations were rationalized on the grounds that the system violates the assumption of measurement independence: the hidden variables, in this case the waveform, depend explicitly on the measurement settings, specifically the system topography. 

While important in framing the question as to how one might violate Bell's inequality with the walking-droplet system, the investigation of Papatryfonos {\it et al.} \cite{PhysRevFluids.9.084001} had {three} principal shortcomings. First, only a static test was possible. As noted by the authors, an extension to a dynamic Bell test would entail isolating the two subsystems before changing the measurement settings. While the particle-particle correlations might survive this isolation, this has not yet been demonstrated. 
Second, the violations were relatively elusive, arising in only a limited corner of parameter space. Third, while the static Bell violations were undoubtedly a manifestation of wave-mediated coupling between the two subsystems, the model is sufficiently complex that the precise dynamical origins of these violations could not be clearly identified. Here, we adopt a relatively simple coupled system of Lorenz-like models that overcomes these three shortcomings. In particular, we achieve relatively robust violations in a static test owing to the enhanced wave coupling in our wave model; moreover, we achieve Bell violations in a dynamic test, and rationalize the emergent violations in terms of wave-mediated chaotic synchronization \cite{Pecora1990}. 

The Lorenz system consists of three coupled ordinary differential equations derived by Edward Lorenz as a low-dimensional approximation to the Navier–Stokes equations governing atmospheric convection~\cite{Lorenz1963}. It has become the canonical model of dissipative chaos {in classical systems} and marked a turning point in our understanding of nonlinear dynamical systems. Lorenz's seminal work revealed that deterministic evolution does not necessarily imply predictability. Instead, chaotic systems exhibit exponential sensitivity to initial conditions, so that arbitrarily small uncertainties in the initial state are amplified, placing a limit on long-term predictability. 
{Notably}, this insight into classical mechanics emerged some three decades after the formulation of the Copenhagen interpretation of quantum mechanics, according to which the quantum state provides a complete description of a quantum system: the possibility of an underlying dynamics is denied. Quantum phenomena are therefore regarded as intrinsically probabilistic. An analogous conclusion in classical mechanics would be that the outcome of a thrown die is intrinsically probabilistic and that its statistical description is complete. We know, however, that the apparent randomness arises from the underlying chaotic dynamics associated with the die interacting with its fluid and solid environments \cite{Kapitaniak2012DieThrow}. 

The field of hydrodynamic quantum analogues \cite{Bush2015,BushOza2020} has revived the question of whether quantum statistics might likewise emerge from an underlying {chaotic} pilot-wave dynamics. Here, we use a Lorenz-like model of interacting walkers to explore the possibility that quantum entanglement is a statistical manifestation of pilot-wave-mediated chaotic synchronization. In \S II, we describe our theoretical model of two vibrating particles trapped in separate subsystems but coupled through their mutual wave field. In \S III, we examine correlations that may arise between the dynamics of the two particles and identify states characterized by both periodic and chaotic synchronization. In \S IV, we identify states that yield static Bell violations. In \S V, we extend our model protocol to consider a classical analog of dynamic Bell tests, in which measurement settings are chosen after the two subsystems are isolated. In \S VI, we identify states that violate Bell's inequality in these dynamic tests, and so exhibit a classical, wave-mediated form of entanglement.

\section{Model} \label{sec:model}

A hierarchy of theoretical models of varying complexity has been developed to describe the walking droplets \cite{BushOza2020}. Some models describe both vertical and horizontal droplet dynamics, as well as the wave dynamics \cite{Molacek2013b, Milewski2015}. Stroboscopic models are deduced by averaging over the timescale associated with wave generation, yielding an equation for the horizontal motion of the droplet, that is seen as a continuous emitter of waves \cite{Molacek2013b,Oza2013}. The resulting droplet trajectory equation is non-Markovian because the wave force acting on the droplet depends on the droplet's history. Specifically, the drop responds to the slope of the local wave field, whose form is deduced by integrating backward in time to account for the waves generated along the droplet's path. 
We here consider reduced models that retain the core nonlinear mechanisms while admitting a finite-dimensional phase-space description. Recent work has shown that, for {specific pilot} waveforms, the integro-differential equations governing a single walking droplet in the stroboscopic model can be transformed exactly into low-dimensional Lorenz-like systems \cite{Durey2020lorenz, ValaniUnsteady, Valanilorenz2022, VALANI2024115253}. This reduction bridges quantum dynamics and chaos theory \cite{Bohm1952a}, recasting the droplet's physical motion in terms of attractors and the structure of phase space.

\begin{figure*}
    \centering
    \includegraphics[width=1.0\columnwidth]{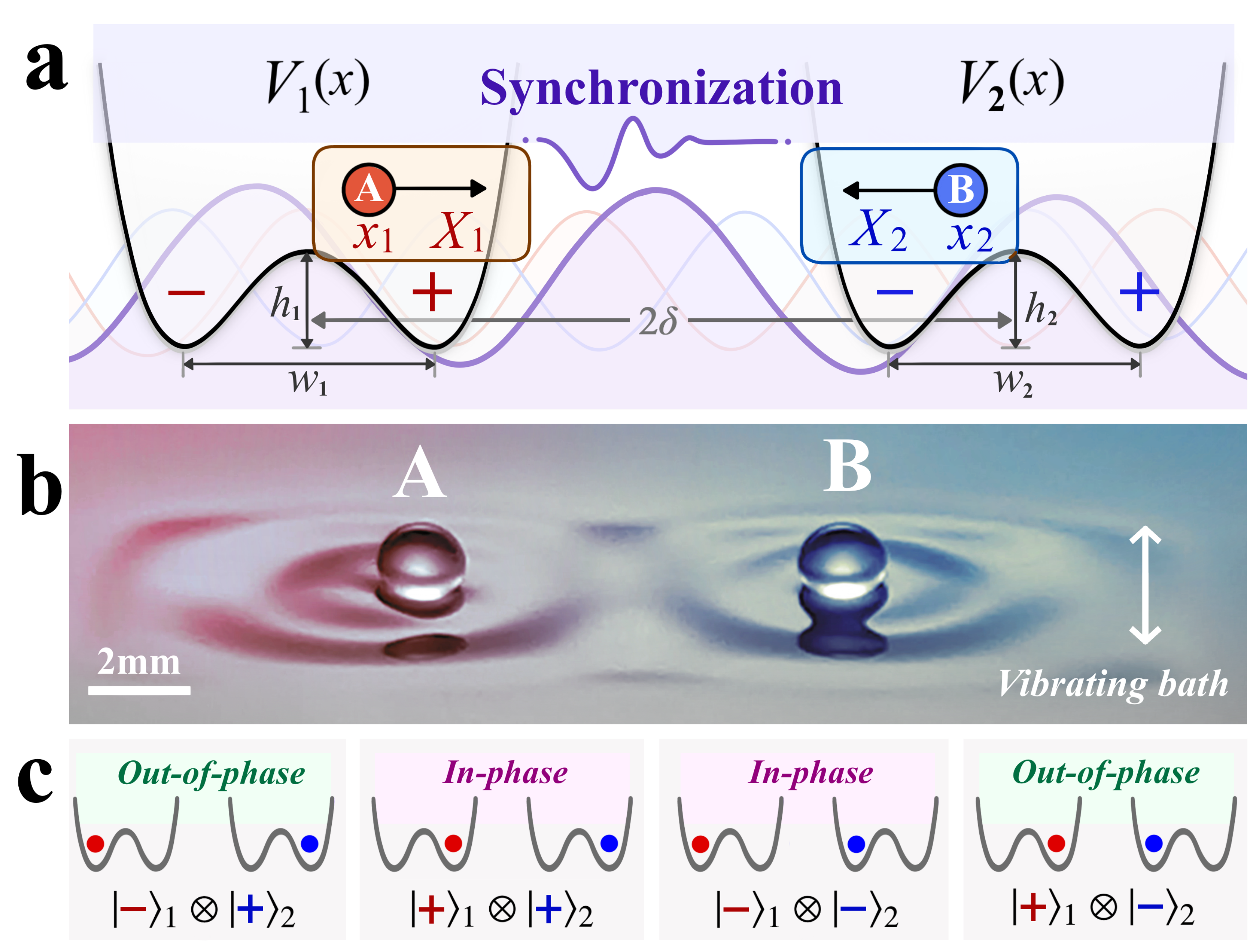}
\caption{Schematic of the coupled two-droplet wave--particle system and its reduced dynamical description. (a) Two droplets, located at positions $x_1$ and $x_2$, move with velocities $X_1$ and $X_2$ in symmetric double-well potentials $V_1(x)$ and $V_2(x)$. Each droplet generates a standing pilot wave with kernel $W(x)=\cos x$, whose amplitude decays exponentially in time. The resulting wave field stores the memory of past impacts and mediates the interaction: each droplet responds to the slope of the combined wave field generated by both droplets. In the reduced Lorenz-like model, these wave-memory effects are represented by auxiliary variables $(Y_{ij}, Z_{ij})$, yielding two coupled low-dimensional subsystems. The double-well geometry is characterized by the well width $w_i$, height $h_i$, and separation $2\delta$. (b) Schematic illustration of the coupled system, showing two walking droplets (A and B) interacting through a common pilot-wave field on a vertically vibrating bath. (c) The four basis states of the bipartite system, defined by the left $|-\rangle$ or right $|+\rangle$ well occupied by each droplet. The in-phase states are $|+\rangle_{1} \otimes |+\rangle_{2}$ and $|-\rangle_{1} \otimes |-\rangle_{2}$, while the out-of-phase states are $|-\rangle_{1} \otimes |+\rangle_{2}$ and $|+\rangle_{1} \otimes |-\rangle_{2}$.}
    \label{fig: schematic}
\end{figure*}

In the present work, we extend this low-dimensional modeling framework to a system of \emph{two interacting droplets} confined to double-well potentials (see Figs.~\ref{fig: schematic}(a) and~\ref{fig: schematic}(b)). We show that the reduced dynamics takes the form of two coupled Lorenz-like subsystems whose symmetry structure admits invariant synchronization manifolds.
Chaotic intermittency and switching between correlated and anticorrelated synchronization states can then generate strong bipartite correlations. Our central aim is to demonstrate that, within this entirely classical and deterministic framework, wave-mediated synchronization and chaos can produce correlations that violate CHSH-type Bell inequalities.
Rather than invoking nonlocality or quantum measurement postulates, we demonstrate that entanglement here arises from the nonlinear dynamics on synchronization manifolds and their intermittent destabilization. This work thus provides a concrete example of how Bell violations may emerge from classical pilot-wave dynamics.

We here adapt the stroboscopic model of Oza \emph{et al.}~\citep{Oza2013}, in which
a droplet is treated as a point particle whose horizontal dynamics are governed by a balance between inertia, drag, and a wave force proportional to the gradient of the local wave field. The horizontal trajectory equation prescribing the evolution of the droplet's horizontal displacement $\mathbf{x}_d(t)$ is then written as~\citep{Molacek2013b,Oza2013}
\begin{equation}
 m\ddot{\mathbf{x}}_d + D\dot{\mathbf{x}}_d
 =
 -mg\nabla h(\mathbf{x}_d,t),
\label{eq:dimensional_particle}
\end{equation}
where $m$ is the droplet mass, $D$ is an effective drag coefficient, and $g$ is the gravitational acceleration. The forcing on the right-hand side arises from the local slope of the self-generated wave field evaluated at the droplet position.

As the droplet bounces on the vertically vibrated bath, it continuously excites standing Faraday waves at each impact. In the stroboscopic model, the resulting surface wave field is modeled as a superposition of standing waves generated along the droplet's path and decaying exponentially in time. The cumulative pilot wave field is given by
\begin{equation}
 h(\mathbf{x},t)
 =
 \frac{A_w}{T_F}
 \int_{-\infty}^{t}
 W\!\left(k_F |\mathbf{x}-\mathbf{x}_d(s)|\right)
 e^{-(t-s)/(T_F\mathrm{Me})}
 \, ds,
\label{eq:dimensional_wavefield}
\end{equation}
where $W(\cdot)$ denotes the spatial structure of the individual standing wave generated by a single impact, $A_w$ is the wave amplitude, $T_F$ is the Faraday period, $k_F=2\pi/\lambda_F$ is the Faraday wavenumber associated with wavelength $\lambda_F$, and $\mathrm{Me}$ is the memory parameter prescribing the temporal decay rate of the waves. In pilot-wave hydrodynamics, quantum-like features typically arise in the high-memory limit, where the pilot wave is most persistent, and the non-Markovian nature of the droplet dynamics is most pronounced.

 Substituting Eq.~\eqref{eq:dimensional_wavefield} into Eq.~\eqref{eq:dimensional_particle} yields the following integro-differential equation governing the horizontal dynamics:
\begin{align}
 m\ddot{\mathbf{x}}_d + D\dot{\mathbf{x}}_d
 =
 &-\frac{mgAk_F}{T_F}
 \int_{-\infty}^{t}
 W'\!\left(k_F|\mathbf{x}_d(t)-\mathbf{x}_d(s)|\right)
 \nonumber \\
 &\times
 \frac{\mathbf{x}_d(t)-\mathbf{x}_d(s)}{|\mathbf{x}_d(t)-\mathbf{x}_d(s)|}
 e^{-(t-s)/(T_F\mathrm{Me})}
 \, ds.
\label{eq:dimensional_IDE}
\end{align}
where $W'$ denotes the derivative of $W$ with respect to its argument. To nondimensionalize the system, we scale space and time according to
\begin{equation}
 \mathbf{\tilde{x}} = k_F\mathbf{x},
 \qquad
 \tilde{t} = \frac{D}{m}t,
\end{equation}
and subsequently drop the tilde for notational convenience. The resulting dimensionless integro-differential equation becomes
\begin{align}
 \ddot{\mathbf{x}}_d + \dot{\mathbf{x}}_d
 =
 -R
 \int_{-\infty}^{t}
 W'\!\left(|\mathbf{x}_d(t)-\mathbf{x}_d(s)|\right)
 \frac{\mathbf{x}_d(t)-\mathbf{x}_d(s)}{|\mathbf{x}_d(t)-\mathbf{x}_d(s)|}
 e^{-(t-s)/\tau}
 \, ds,
\label{eq:dimensionless_2D}
\end{align}
where the dimensionless parameters are
\begin{equation}
 R = \frac{m^3gAk_F^2}{D^3T_F},
 \qquad
 \tau = \frac{DT_F\mathrm{Me}}{m},
\end{equation}
represent, respectively, the dimensionless wave forcing amplitude and the dimensionless memory time.

In the present study, we focus on a reduced one-dimensional geometry in which the droplets are constrained to move along a horizontal line. Such one-dimensional reductions have been widely employed in theoretical studies of walking droplets and retain the essential ingredients responsible for memory-driven instabilities and nonlinear dynamics~\citep{durey2018, Durey2020lorenz, ValaniUnsteady, phdthesismolacek, Valanilorenz2022}. The corresponding dimensionless equation of motion for a single droplet (we drop the subscript `d' for convenience) becomes
\begin{equation}
 \ddot{x} + \dot{x}
 =
 -R
 \int_{-\infty}^{t}
 W'\!\left(x(t)-x(s)\right)
 e^{-(t-s)/\tau}
 \, ds.
\label{eq:1D_single_droplet}
\end{equation}
The simple form taken for the pilot-wave $W(x)= \cos{x}$ will be justified in \S II.B.

In the walking droplet system, each impact generates a circularly symmetric wave that propagates radially outward at approximately 23 cm/s, behind which persists a quasi-monochromatic stationary wave with Bessel form, $J_0(k_F r)$~\cite{Eddi2011a, Molacek2013b}. In the stroboscopic model ~\cite{Oza2013}, the propagating front is neglected, so a temporally decaying Bessel wave form is added to the wave field at each point along the droplet's path. Although the stroboscopic model has been successful in rationalizing the dynamics of free walkers and walkers moving in response to applied forces~\cite{Oza2014a, Perrard2014, Labousse2014}, it has shortcomings in capturing the detailed dynamics of interacting droplet pairs in experiments ~\cite{Arbelaiz2018, Oza2018, Couchman2019}. The stroboscopic model was found to be particularly inadequate at capturing the behavior of ratcheting pairs, whose dynamics is dominated by both variability in the vertical dynamics and the influence of the propagating front \cite{GaleanoRios2018}. More accurate modeling of walker-walker interactions has generally required consideration of a more complete description of the pilot-wave that captures its finite speed of propagation ~\cite{Milewski2015, correlationnachbin}. In the context of our study, we adopt the stroboscopic model not as a faithful representation of a particular experimental system but as a convenient theoretical 
framework {that captures the essential physics of pilot-wave hydrodynamics.}

\subsection{Two interacting droplets in double-well potentials}

We now extend the model in order to describe two interacting droplets with horizontal positions $x_1(t)$ and $x_2(t)$. Each droplet is confined to lie within an external double-well potential and generates waves that influence both the droplet {and its partner} through the underlying wave field. The equation of motion for droplet $i$ takes the form
\begin{equation}
 \ddot{x}_i + \dot{x}_i
 =
 F_i^{\mathrm{self}}[x_i]
 +
 F_i^{\mathrm{int}}[x_i,x_j]
 +
 F_i^{\mathrm{ext}}[x_i],
\label{eq:two_droplet_general}
\end{equation}
with $i\neq j$, $F_i^{\mathrm{self}}$ denotes the wave force arising from the droplet's own past trajectory, $F_i^{\mathrm{int}}$ denotes the interaction force arising from the wave field generated by the other droplet, and $F_i^{\mathrm{ext}}$ represents the force produced by the external confining potential. The self-memory and interaction-memory forces are given explicitly by
\begin{align}
 F_i^{\mathrm{self}}(t)
 &=
 -R
 \int_{-\infty}^{t}
 W'\!\left(x_i(t)-x_i(s)\right)
 e^{-(t-s)/\tau}
 \, ds,
 \\
 F_i^{\mathrm{int}}(t)
 &=
 -R
 \int_{-\infty}^{t}
 W'\!\left(x_i(t)-x_j(s)\right)
 e^{-(t-s)/\tau}
 \, ds.
\end{align}
{The external confinement is described by a piecewise double-well
potential,
\begin{equation}
V(x)=
\begin{cases}
\dfrac{1}{4}A_1(x+\delta)^4
-\dfrac{1}{2}B_1(x+\delta)^2, & x<0,\\[2mm]
\dfrac{1}{4}A_2(x-\delta)^4
-\dfrac{1}{2}B_2(x-\delta)^2, & x\geq 0,
\end{cases}
\label{eq:double_well_potential}
\end{equation}
obtained by joining two double-well potentials centered at
$x=\pm\delta$. The parameter $2\delta$ thus specifies the separation
between their centers. Each droplet is confined to its corresponding
half of the potential, so that
\begin{equation}
V_1(x)=V(x)\big|_{x<0},
\qquad
V_2(x)=V(x)\big|_{x\geq0}.
\end{equation}
The corresponding external forces are
\begin{equation}
F_i^{\mathrm{ext}}(x_i)=-V_i'(x_i),
\end{equation}
or explicitly,
\begin{align}
F^{\mathrm{ext}}_1(x_1)
&=-A_1(x_1+\delta)^3+B_1(x_1+\delta),\\
F^{\mathrm{ext}}_2(x_2)
&=-A_2(x_2-\delta)^3+B_2(x_2-\delta).
\end{align}
}
For the most general simulations presented in this work, we consider different double-well potentials with respective parameters $(A_1, B_1)$ and $(A_2, B_2)$. The parameters $A_i$ and $B_i$ are related to the width $w_i$ and height $h_i$ of the double-well potential according to
\begin{equation}
 A_i=\frac{64h_i}{w_i^4},
 \qquad
 B_i=\frac{16h_i}{w_i^2}.
\end{equation}
Unless otherwise specified, we shall consider that both double-wells have the same width $w_1=w_2=1.5\lambda_F$, where the Faraday wavelength corresponds to $\lambda_F=2\pi$. Similarly, we fix the distance between the maxima of the two wells to be $2 \delta=7\lambda_F$. Since each droplet is restricted to its respective half-domain, the piecewise potential need not be continuous at $x=0$; any discontinuity there has no influence on the droplet dynamics.

\subsection{Reduction to a coupled Lorenz-like system}

The essential feature of the walking-droplet wave field is its oscillatory structure with the Faraday wavelength. While experimentally observed wave profiles are accurately described by spatially damped Bessel functions~\citep{Damiano2016,Couchman2019}, previous studies have shown that many of the key qualitative dynamical features of one-dimensional pilot-wave dynamics can be captured using the simplified sinusoidal kernel~\citep{Durey2020lorenz, ValaniUnsteady, phdthesismolacek, Valanilorenz2022}
\begin{equation}
 W(x)=\cos(x),
 \qquad
 W'(x)=-\sin(x).
\end{equation}
This idealized choice retains the oscillatory character of the wave-mediated interaction while allowing an exact reduction of the integro-differential equation to a finite-dimensional system of ordinary differential equations. {Moreover, this choice ensures maximal coupling between the two subsystems.} 

Following previous work on reduced walking-droplet models~\citep{Durey2020lorenz, Valanilorenz2022,Valani2024}, we now introduce auxiliary memory variables that encode exponentially weighted wave-history integrals and thereby transform the system into a finite-dimensional set of ordinary differential equations. We define the auxiliary {memory variables as follows:}
\begin{align}
 Y_{ij}(t)
 &=
 R
 \int_{-\infty}^{t}
 \sin\!\left(x_i(t)-x_j(s)\right)\,e^{-(t-s)/\tau}
 \, ds,
 \label{eq:memory_variables_1}
 \end{align}
 \begin{align}
 Z_{ij}(t)
 &=
 R
 \int_{-\infty}^{t}
 \cos\!\left(x_i(t)-x_j(s)\right)\,e^{-(t-s)/\tau}
 \, ds.
\label{eq:memory_variables}
\end{align}
The variables $Y_{ij}$ encode the wave-mediated memory force on droplet $i$ arising from the past trajectory of droplet $j$, while $Z_{ij}$ encode the corresponding wave amplitudes at the droplet location. Owing to the time-delayed nature of the interaction, the cross-memory terms are generally non-reciprocal~\cite{refId0,Bowick2022,Gupta2022},
$Y_{ij}\neq -Y_{ji}$. Differentiating Eqs.~\eqref{eq:memory_variables_1}-\eqref{eq:memory_variables} with respect to time and applying Leibniz's rule yields
\begin{align}
 \dot{Y}_{ij}
 &=
 -\frac{1}{\tau}Y_{ij}
 +
 \dot{x}_i Z_{ij}
 +
 R\sin(x_i-x_j),
 \\
 \dot{Z}_{ij}
 &=
 -\frac{1}{\tau}Z_{ij}
 -
 \dot{x}_i Y_{ij}
 +
 R\cos(x_i-x_j).
\end{align}
For the {self-memory terms} with $i=j$, these equations reduce to
\begin{align}
 \dot{Y}_{ii}
 &=
 -\frac{1}{\tau}Y_{ii}
 +
 \dot{x}_i Z_{ii},
 \\
 \dot{Z}_{ii}
 &=
 R
 -
 \frac{1}{\tau}Z_{ii}
 -
 \dot{x}_i Y_{ii},
\end{align}
since $\sin(0)=0$ and $\cos(0)=1$. Introducing the velocity variables
\begin{equation}
 X_1=\dot{x}_1,
 \qquad
 X_2=\dot{x}_2,
\end{equation}
and the parameter $\zeta=\{0,1\}$, which allows us to switch off the coupling between droplets, we can write the twelve-dimensional system of differential equations. The coupled Lorenz-like dynamical system governing droplet~1 is
\begin{align}
 \dot{x}_1
 &=
 X_1,
 \nonumber \\
 \dot{X}_1
 &=
 -X_1
 +
 Y_{11}
 +
 \zeta Y_{12}
 +
 F_1(x_1),
 \nonumber \\
 \dot{Y}_{11}
 &=
 -\frac{1}{\tau}Y_{11}
 +
 X_1Z_{11},
 \nonumber \\
 \dot{Y}_{12}
 &=
 -\frac{1}{\tau}Y_{12}
 +
 X_1Z_{12}
 +
 R\zeta\sin(x_1-x_2),
 \nonumber \\
 \dot{Z}_{11}
 &=
 R
 -
 \frac{1}{\tau}Z_{11}
 -
 X_1Y_{11},
 \nonumber \\
 \dot{Z}_{12}
 &=
 -
 \frac{1}{\tau}Z_{12}
 -
 X_1Y_{12}
 +
  R\zeta\cos(x_1-x_2).
\label{eq:droplet1_lorenz}
\end{align}

The corresponding equations for droplet~2 are
\begin{align}
 \dot{x}_2
 &=
 X_2,
 \nonumber \\
 \dot{X}_2
 &=
 -X_2
 +
 Y_{22}
 +
 \zeta Y_{21}
 +
 F_2(x_2),
 \nonumber \\
 \dot{Y}_{22}
 &=
 -\frac{1}{\tau}Y_{22}
 +
 X_2Z_{22},
 \nonumber \\
 \dot{Y}_{21}
 &=
 -\frac{1}{\tau}Y_{21}
 +
 X_2Z_{21}
 +
 R\zeta\sin(x_2-x_1),
 \nonumber \\
 \dot{Z}_{22}
 &=
 R
 -
 \frac{1}{\tau}Z_{22}
 -
 X_2Y_{22},
 \nonumber \\
 \dot{Z}_{21}
 &=
 -
 \frac{1}{\tau}Z_{21}
 -
 X_2Y_{21}
 + R\zeta\cos(x_2-x_1).
\label{eq:droplet2_lorenz}
\end{align}

The resulting formulation consists of two coupled Lorenz-like subsystems interacting through nonlinear sine and cosine coupling terms. The {memory} variables $(Y_{ij}, Z_{ij})$ encode the memory resulting from the wave-mediated interaction between the droplets and render the original integro-differential dynamics finite-dimensional. This reduction provides a transparent framework for analyzing synchronization manifolds, chaotic intermittency, and the emergence of strong bipartite correlations in the walking-droplet system. We solve the coupled dynamical system in Eqs.~\eqref{eq:droplet1_lorenz} and \eqref{eq:droplet2_lorenz} using MATLAB solver ode45. {Unless otherwise stated, the initial droplet positions are sampled
uniformly across their respective double wells, the velocities are
sampled uniformly from $[-0.01,0.01]$, and all wave-memory variables
$Y_{ij}$ and $Z_{ij}$ are} {initially set to zero, corresponding 
to a flat interface.}

\section{Entanglement as synchronization} \label{ent:sync}

Entanglement is conceptualized here as a synchronization phenomenon between two droplets mediated by the underlying wave field that serves as the system's memory. More generally, the correlated motion observed in our system can be understood within the framework of synchronization theory for coupled nonlinear dynamical systems. Synchronization arises when interactions constrain the dynamics of coupled subsystems, causing them to evolve in invariant subspaces of the full phase space known as synchronization manifolds~\cite{pikovsky1985universal, StrogatzBook,Strogatz2000FromKT}.
\begin{figure*}
    \centering
    \includegraphics[width=1\columnwidth]{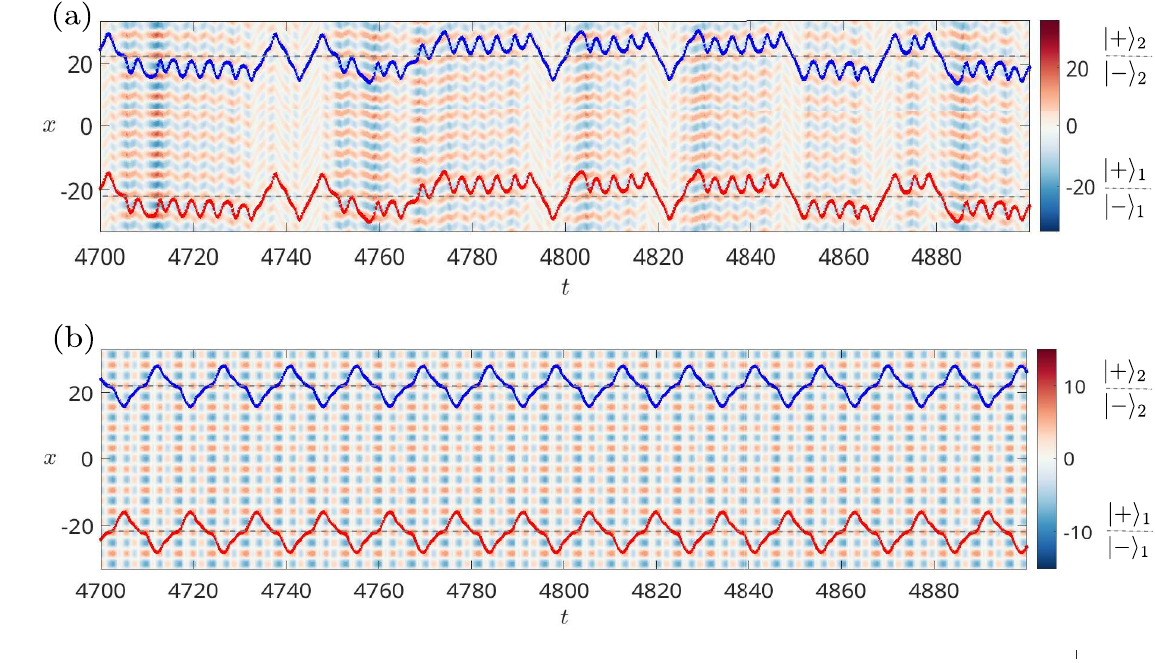}
  \caption{Correlated dynamics of two walking droplets in the two synchronization manifolds. The time series of the positions of the walking droplets 1 and 2 are depicted in red and blue, respectively. In the background, the instantaneous wave height $h(x,t)$ is plotted  (see color bar). (a) Two droplets showing synchronized chaotic intermittency~\citep{Cvitanovic2020Chaos} in the correlated manifold $S_1$, for $h_1=h_2=6.0$, $\tau=6.8$, and $R=4.4$. The droplets are almost perfectly synchronized with synchronization error $\xi(t)=\norm{(x_1(t)+\delta)-(x_2(t)-\delta)}<10^{-10}$. The von Neumann entropy of the resulting state \( |\psi\rangle = (|-\rangle_{1} \otimes |-\rangle_{2} + |+\rangle_{1} \otimes |+\rangle_{2})/\sqrt{2} \) is $S_{vN}=0.99$, typical of maximally entangled states. (b) Two droplets showing synchronized periodic oscillations in the anticorrelated manifold $S_2$, for $h_1=h_2=3.5$, $\tau=0.85$, and $R=5.0$. The droplets are almost perfectly synchronized with synchronization error $\xi(t)=\norm{(x_1(t)+\delta)+(x_2(t)-\delta)}<10^{-10}$. The von Neumann entropy of the resulting state  \( |\psi\rangle = (|-\rangle_{1} \otimes |+\rangle_{2} + |+\rangle_{1} \otimes |-\rangle_{2})/\sqrt{2} \) is $S_{vN}=1.0$, typical of maximally entangled states. See also Supplemental Video S1.}
 \label{fig: wavefield sync}
\end{figure*}

Memory provides the key dynamical ingredient required for synchronization in the present system. It is well established that memory-based dynamical systems may undergo Hopf bifurcations and develop self-sustained oscillations~\cite{Lopez2020, LOPEZ2023113412}, naturally connecting the present wave-driven particles to the broader class of active particles characterized by autonomous, self-sustained dynamics~\cite{PismenActiveMatterBook, VALANI2024115253}. In pilot-wave hydrodynamics, such oscillations between droplet pairs may arise via coupling through the pilot wave, giving rise to distributed time-delay interactions between the droplets.
Synchronization phenomena arising from such wave-mediated interactions have been extensively investigated in pilot-wave hydrodynamics~\cite{Nachbin2018, Saenz2018b, Thomson2019, Couchman2020,Nachbin2022,CorrNachbin2026SpontaneousOW}. 

Representative trajectories illustrating the emergence of synchronization and strong correlations in our system are shown in Fig.~\ref{fig: wavefield sync}. In the parameter regime characterized by strong coupling and chaotic switching between the wells, the observed correlations emerge from the synchronization of chaotic intermittency~\cite{VALANI2024115253} (see Fig.~\ref{fig: wavefield sync}(a)). Here, the individual trajectories remain chaotic and therefore intrinsically unpredictable, yet the wave-mediated coupling constrains the dynamics to a synchronization manifold on which switching events occur in a correlated fashion. Consequently, two seemingly contradictory dynamical features, strong correlations and intrinsic unpredictability, can coexist within pilot-wave hydrodynamics.

Strong correlations also arise in synchronized periodic regimes. In this case, both droplets evolve on stable limit cycles and switch coherently between the wells of their respective double-well potentials (Fig.~\ref{fig: wavefield sync}(b)). Thus, both chaotic and periodic synchronization provide dynamical routes to the strong correlations explored in this work.

\subsection{Measuring entanglement}
 
Specific measures of entanglement and synchronization can be introduced by associating each well with a corresponding eigenstate~\cite{PhysRevFluids.9.084001}. Within this framework, bipartite states \( |\psi\rangle \) are constructed as superpositions of the four elementary tensor-product states \(|\pm\rangle_{1}\otimes|\pm\rangle_{2} \). Each of these basic states represents the location of the particles in one of the four possible double-well combinations, as shown in Fig.~\ref{fig: schematic}(c). The notion of statistical superpositions being rooted in intermittent switching between unstable, quantized orbital states \cite{MegastableLopezValani2025} has been extensively investigated in prior studies of orbital pilot-wave hydrodynamics~\cite{Harris2013, Oza2014, Perrard2014, Labousse2014a,BushOza2020}, and is adopted here. 
In certain instances, the emergent steady statistics are related to the mean pilot-wave field \cite{Saenz2018, durey2018, bush2024perspectives}.

The coefficients \(a_{\pm,\pm}\) that define the resulting pure bipartite state \( |\psi\rangle \) are computed from the mean fraction of time \(\tau_{\pm\pm}\) spent by the system in each of the four possible well configurations over sufficiently long trajectories, 
that is,
\begin{equation}
a_{\pm \pm} =  \lim_{t\rightarrow\infty}\sqrt{\dfrac{\tau_{\pm \pm}(t)}{t}}.
\end{equation} 

From the resulting pure state \( |\psi\rangle \), the reduced density matrix $\rho_{1}=\mathrm{tr}_{2}\!\left(|\psi\rangle\langle\psi|\right)$ associated with one of the two droplets (here chosen, without loss of generality, as droplet \(1\)) can be readily computed \footnote{Explicitly, the pure state can be written as \( |\psi\rangle =a_{++}|+\rangle_1\otimes|+\rangle_2+a_{+-}|+\rangle_1\otimes|-\rangle_2+a_{-+}|-\rangle_1\otimes|+\rangle_2 + a_{--}|-\rangle_1\otimes|-\rangle_2 \). Therefore, the reduced density matrix can be calculated as $(\rho_1)_{bc}=\sum_d a_{b d}a_{c d}^{*}$. The two eigenvalues $\lambda_a$ of the two-dimensional matrix $\rho_1$ allow to compute the von Neumann entropy as $S_{vN}=-\sum_{a}\lambda_a\log(\lambda_a)$.}. The corresponding von Neumann entropy,
\begin{equation}
S_{vN}(\rho_{1})=-\mathrm{tr}\!\left(\rho_{1}\log\rho_{1}\right),
\end{equation}
provides a standard measure of entanglement for the synchronized bipartite state  \cite{Horodecki2009}. In particular, \(S_{vN}=1\) corresponds to a maximally entangled state, whereas \(S_{vN}=0\) characterizes a separable, non-entangled state.

\subsection{Synchronization manifolds}

Consider two identical dynamical systems coupled symmetrically through an interaction function as per below
\begin{align}\label{eq: general DS}
\dot{\mathbf{r}}_1 &= F(\mathbf{r}_1) + H(\mathbf{r}_1-\mathbf{r}_2),\\ 
\nonumber
\dot{\mathbf{r}}_2 &= F(\mathbf{r}_2) + H(\mathbf{r}_2-\mathbf{r}_1),
\end{align}
where $\mathbf{r}_1,\mathbf{r}_2 \in \mathbb{R}^n$, the vector field function $F:\mathbb{R}^n \rightarrow \mathbb{R}^n$ describes the dynamics of each subsystem, and the vector field {function} $H:\mathbb{R}^n \rightarrow \mathbb{R}^n$ captures the coupling between them. 

In general, a \emph{synchronization manifold} is an invariant subspace of the full phase space \cite{eroglu2017synchronisation} on which the trajectories of the two subsystems satisfy a prescribed functional relation
\[
G(\mathbf{r}_1,\mathbf{r}_2)=0.
\]
This mathematical condition is frequently used to assess the degree of synchronization \cite{eroglu2017synchronisation} by defining the synchronization error $\xi(t)=\norm{G(\mathbf{r}_1,\mathbf{r}_2)}$. If the dynamics is initialized on such a manifold, it remains confined in it for all subsequent times, yielding the value $\xi=0$. Moreover, whenever any of these synchronization subspaces is globally stable, trajectories originating from arbitrary initial conditions converge asymptotically towards the synchronized state.

The simplest and most widely studied example is the complete synchronization manifold \(\mathbf{r}_1=\mathbf{r}_2\), which corresponds to perfectly correlated motion between the two subsystems \cite{eroglu2017synchronisation}. Furthermore, for certain synchronization manifolds, such as the complete synchronization manifold in diffusively coupled systems, the interaction term identically vanishes on the manifold
\[
H(\mathbf{r}_1,\mathbf{r}_2)\big|_{G(\mathbf{r}_1,\mathbf{r}_2)=0}=0,
\]
implying that synchronization can \emph{persist} even after decoupling the two particles, so that they may remain correlated arbitrarily long times in the absence of further interaction \cite{eroglu2017synchronisation}.

\subsection{Symmetry groups}

A fundamental question concerns the number of synchronization spaces in a dynamical system {characterized by a vector field function $F$, and the stability of these spaces}. In the Appendix~\ref{sec:A2}, we show that, under appropriate conditions on the coupling function $H$, each symmetry of the underlying dynamical system $F$ generates a corresponding symmetry-related synchronization manifold. In particular, the coupled Lorenz-like droplet model exhibits spatial reflection symmetry due to the symmetric placement of the double-well potentials about the origin. Specifically, the single-droplet dynamics are invariant under the reflection transformation $(x\pm \delta, X, Y, Z) \rightarrow (-(x\pm \delta),-X,-Y, Z)$~\footnote{In particular, the system is invariant under the discrete two-element reflection group, $i.e.$ the cyclic group $C_2=\{e,g\}$, where $e$ denotes the identity transformation and $g$ corresponds to a reflection about the midpoint between the two wells. A faithful representation of this group in our Lorenz-like system is given by $D: G \rightarrow GL(4,\mathbb{R})$, where the matrices are given by $D(e)=\text{diag}(1,1,1,1)$ and $D(g)=\text{diag}(-1,-1,-1,1)$.}.
\begin{figure*}
    \centering
    \includegraphics[width=1.0\columnwidth]{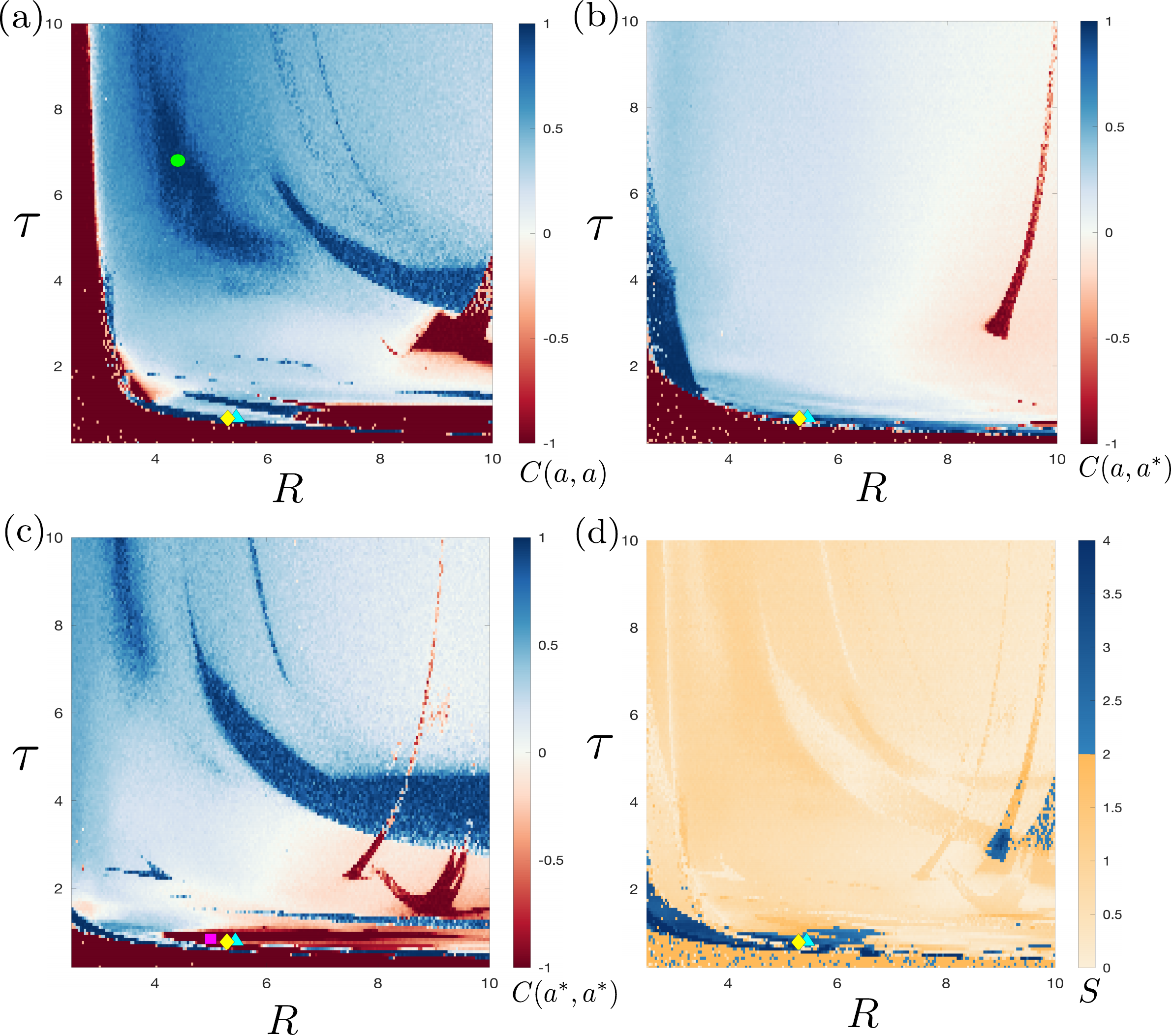}
    \caption{Dependence of the two-particle correlations on the memory parameter $\tau$ and the wave forcing parameter $R$. Colormaps of (a) $C(a,a)$, (b) $C(a,a^{*})$, (c) $C(a^{*},a^{*})$, and (d) the Bell parameter, $|S|=|C(a,a) - C(a^{*},a^{*}) + 2\,C(a,a^{*})|$, shown as functions of $(R,\tau)$. In panel (d), Bell inequality violations ($|S|>2$) are highlighted in blue. The measurement settings $(a,a)$ and $(a^{*},a^{*})$ correspond to $h_1=h_2=6$ and $h_1^*=h_2^*=3.5$, respectively. The markers indicate parameter values examined in other figures: the green circle corresponds to Fig.~\ref{fig: wavefield sync}(a), the magenta square to Fig.~\ref{fig: wavefield sync}(b), the cyan triangle to Fig.~\ref{fig: traj 3}, and the yellow diamond to the interaction regime shown in Fig.~\ref{fig: S4}.}
    \label{fig: 1 in phase}
\end{figure*}

This symmetry gives rise to the two dynamically relevant invariant manifolds $S_1$ and $S_2$, corresponding to the correlated (trivial manifold) and anticorrelated droplet motion, respectively. As shown in the Appendix~\ref{sec:A2}, these manifolds are invariant because the intrinsic dynamics in Eq.~\eqref{eq: general DS} preserve the corresponding symmetry transformations, while the coupling function $H$ satisfies the associated compatibility
conditions.
\begin{figure*}
    \centering
    \includegraphics[width=1\columnwidth]{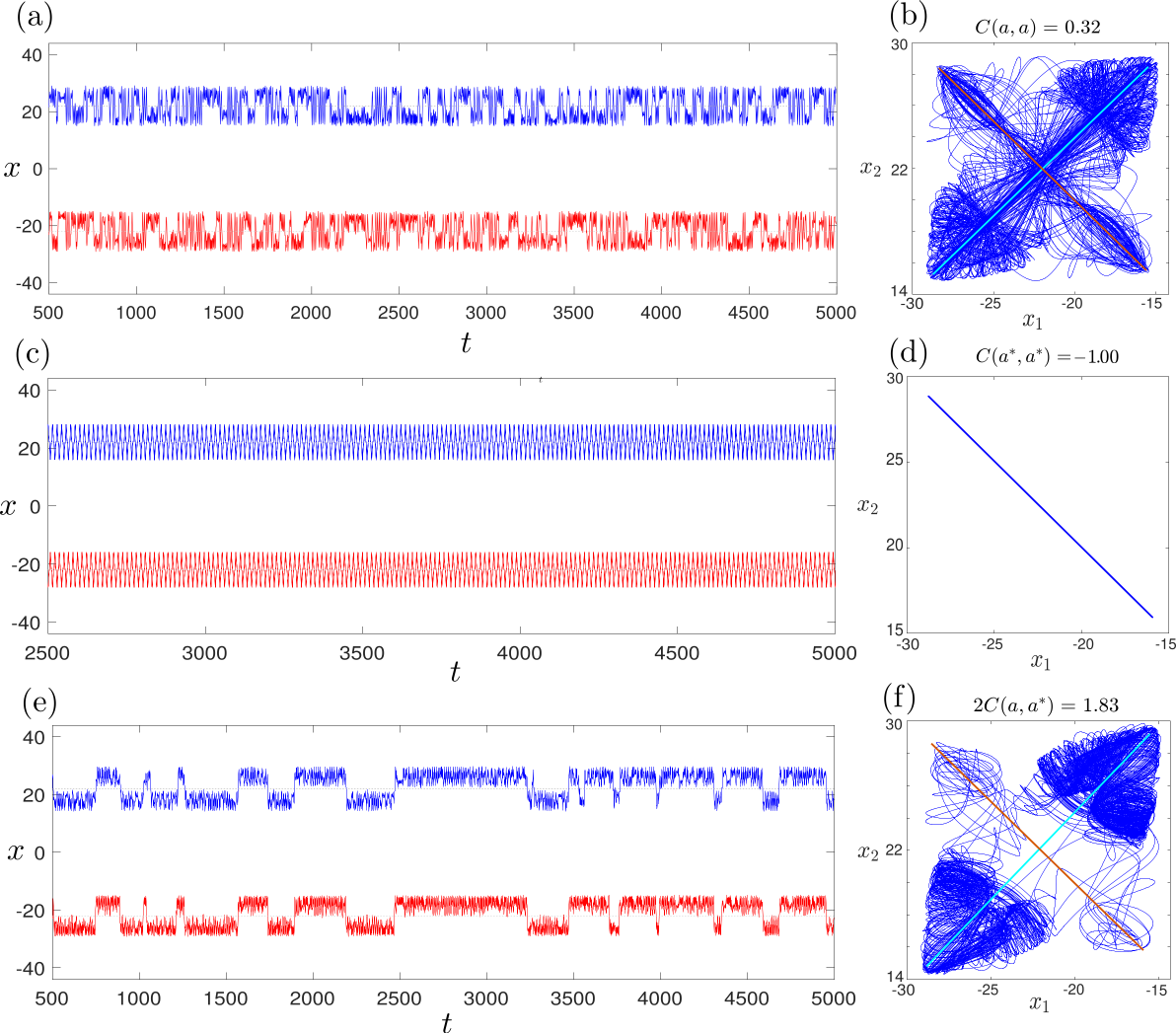}
    \caption{Trajectories {responsible} for static Bell violations with $S\approx3.15$ for two walking droplets at $R=5.44$ and $\tau=0.8$. (a) Setting $(a, a)$ corresponds to $h_1=h_2=6$, yielding chaotic intermittent motion between (b) synchronization manifolds, which are both unstable, with a slight dominance of $S_1$ (light blue) over $S_2$ (light orange), yielding a small positive correlation. (c) Setting $(a^{*},a^{*})$ to $h_1^*=h_2^*=3.5$ yields periodic, perfectly anticorrelated dynamics, since the (d) synchronization manifold $S_2$ is stable. (e) Setting $(a, a^{*})$ corresponds to $h_1=6.0$, $h_2^*=3.5$, and yields a chaotic intermittent motion between (f) synchronization manifolds, which are both unstable, with greater dominance of $S_1$ (light blue) over $S_2$ (light orange), yielding a strong positive correlation. See also Supplemental Video S2.}
    \label{fig: traj 3}
\end{figure*}

In our Lorenz-like system, the first manifold is
\[
S_1
=
\left\{
(x_1,x_2)\in\mathbb{R}^{2}
\,:\,
x_1+\delta=x_2-\delta
\right\},
\]
corresponds to a \emph{correlated} synchronization state. As shown in Fig.~\ref{fig: wavefield sync}(a), in this manifold, the positions of the two droplets are identical with respect to the centers of their respective double-wells. Equivalently, the droplets occupy the corresponding states in each well and evolve synchronously in phase with equal relative displacements from the local minima. The second manifold
\[
S_2 = \left\{
(x_1,x_2)\in\mathbb{R}^2
\,:\,
x_1+\delta=-(x_2-\delta)
\right\},
\]
corresponds to an \emph{anticorrelated} synchronization state. As shown in Fig.~\ref{fig: wavefield sync}(b), in this manifold, the displacement of one droplet from the center of its well is equal in magnitude but opposite in direction to that of the other droplet. The two droplets, therefore, occupy opposite states relative to the center of their respective wells, {and execute} mirror-image trajectories about $x=0$.

Depending on the global stability of these manifolds (see Appendix~\ref{sec:A3}), the particles can be attracted to these synchronization spaces or not. Consequently, trajectories may lead asymptotically to {either} of these manifolds, each having a certain basin of attraction. When any of these manifolds is stable, the asymptotic dynamics evolve towards a maximally entangled state. 
Where the two manifolds are unstable, trajectories can evolve close to either synchronization manifold for long periods of time before intermittently transitioning between them due to chaotic instability. This introduces a new scale for the intermittency of the dynamics, since the synchronization of the droplets' intermittency can itself be intermittent on a longer time scale, resulting in a \emph{multiscale} critical system. 

As we show below in our numerical simulations, this switching between correlated and anticorrelated synchronization states plays a central role in the emergence of strong bipartite correlations in some settings of our {dynamical system}. In particular, the long-time statistics associated with trajectories evolving near $S_1$ and $S_2$ determine the sign and magnitude of the measured correlation functions that enter the Bell inequalities {and ultimately lead} to violations. Simply put, the average transient time spent close to each of these subspaces determines the degree of entanglement, as quantified by the von Neumann entropy.

\section{Bell correlations and the Violation Criterion}

No-go theorems are intended to constrain the type of classical hidden-variable models that might plausibly provide the dynamical underpinnings of the theory of quantum statistics \cite{Brukner2012Bell}. In our Lorenz-like droplet model, the hidden variables are the position and momentum of each droplet $(x_{i}, X_{i})$, and the wave field $h(\textbf{x},t)$ from which the {memory} forces $Y_{i j}$ and the wave height $Z_{i j}$ are computed at the droplet positions. 
{Notably,} these wave variables extend throughout the fluid domain, and so are \emph{not} restricted 
to particles, thus representing extrinsic rather than intrinsic variables \cite{Vervoort2018}.

Consider two measurement devices, each characterized by one of two possible measurement settings, denoted $(a,a^{*})$ for the first subsystem and $(b,b^{*})$ for the second. The bipartite correlation $C(a,b)$ denotes the correlation between the corresponding measurement outcomes obtained for the pair of settings $(a,b)$. According to Bell~\cite{Bell1964}, if one assumes a probability density $p(\lambda)$ for the hidden variables $\lambda$ of the bipartite system, then these correlations can be expressed as
\begin{equation}
C(a,b)=\int_{\Lambda}A_{a}(\lambda)B_{b}(\lambda) p(\lambda)\,d \lambda,
\label{eq:28}
\end{equation}
where the functions $A_{a}(\lambda)$ and $B_{a}(\lambda)$ assign a value of $\pm 1$ to the hidden variables through the measurement process. 

In the CHSH-Bell formulation \cite{clauser1970proposed}, the Bell parameter \(S\) is defined in terms of the bipartite correlations \(C(a,b)\) as
\begin{equation}
S = C(a,b) + C(a^{*},b) + C(a,b^{*}) - C(a^{*},b^{*}),
\label{eq:30}
\end{equation}
with the Bell inequality being
\begin{equation}
|S| \le 2.
\label{eq:31}
\end{equation}
If Eqs.~\eqref{eq:28},~\eqref{eq:30} are satisfied by a hidden-variable theory, then it is widely believed that violation of Eq.~\eqref{eq:31} requires that theory to be non-local \cite{Maudlin2014}. Nevertheless, robust violations of Eq.~\eqref{eq:31} have been reported {in quantum systems} \cite{Aspect1981, Aspect1982b}. We note further that in quantum mechanics, bipartite states of photons must satisfy the Tsirelson bound $S \leq 2\sqrt{2}$ \cite{larsson2014loopholes}. 

The independence of the probability density $p(\lambda)$ on the measurement settings of the apparatus $(a,b)$ is known as the condition of {\it measurement independence} and can be formally written as $p(\lambda|a,b)=p(\lambda)$. The relevance of this condition for pilot-wave systems has been questioned by Vervoort~\cite{Vervoort2018}. Static Bell tests with walking droplets have been shown to violate Eq.~\eqref{eq:31}, {and these Bell violations were rationalized on the basis of the system not complying} with the assumption of measurement independence \cite{PhysRevFluids.9.084001}.
Here, we corroborate these results in our idealized model of classical pilot-wave dynamics, and detail the dynamical origins of the Bell violations. We further demonstrate that the Bell violations persist even after the wave-mediated communication between the two subsystems is eliminated, thereby achieving the first Bell violation in a dynamic test with a classical pilot-wave system.
\begin{figure}
    \centering
    \includegraphics[width=0.75\columnwidth]{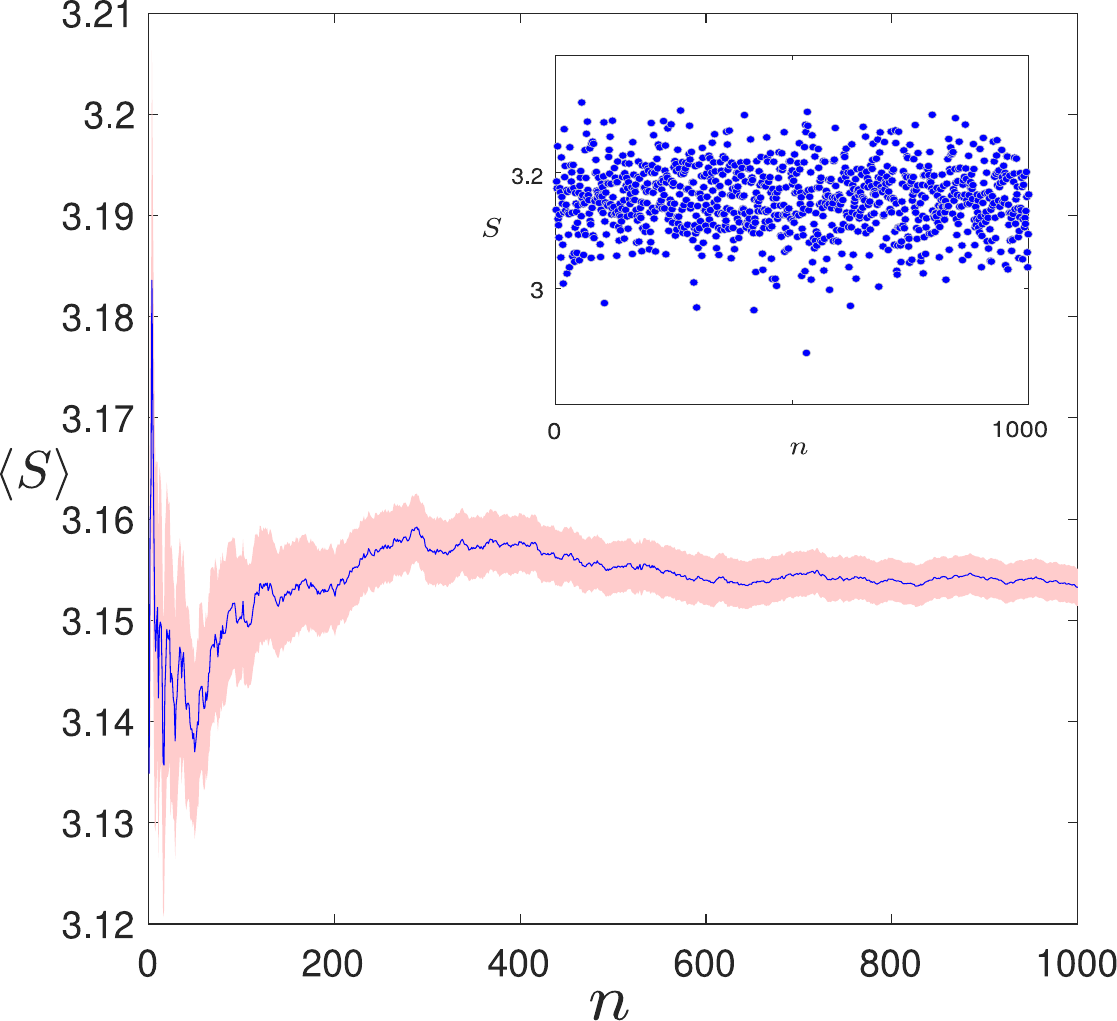}
    \caption{Convergence of the ensemble-averaged CHSH Bell parameter $\langle S\rangle$ with the number of statistically independent initial conditions, $n$, sampled uniformly over the double-well phase space. The integration time was $t = 10^4$. The mean converges to $\langle S\rangle \approx 3.15$, well above the Tsirelson bound, while the shaded region denotes one standard error of the mean. As expected from the law of large numbers, the standard error decreases with increasing sample size, reaching approximately $\sigma_{\langle S \rangle} \approx 1.5\times10^{-3}$ for $n=1000$. The inset shows the corresponding values of $S$ obtained from different initial conditions.}
    \label{fig: violation}
\end{figure}

Deterministic classical field theories have been shown to violate measurement independence, both through general theoretical arguments \cite{quantum7030029, Vervoort2018} and explicit examples \cite{l3xp-yrrv}. From this perspective, the violation of measurement independence in classical field theories points to the role of pre-existing correlations in the hidden background fields \cite{Morgan_2006, Lopez_2023}. In the present case, the deterministic field evolution ensures that the pilot-wave field will depend on the measurement setting through its influence on the droplet motion and so the local geometry of the wave field.

In the symmetric configuration considered here, where $a=b$ and $a^{*}=b^{*}$, the four measurement settings correspond to $(a,a)$, $(a^{*},a^{*})$, $(a,a^{*})$ and $(a^{*},a)$, and the Bell parameter can be expressed as
\begin{equation}\label{eq:S_symmetric}
S = C(a,a) - C(a^{*},a^{*}) + 2\,C(a,a^{*}).
\end{equation}
We have assumed that $C(a, a^{*})=C(a^{*}, a)$, since we use initial conditions that are uniformly distributed over the two wells. (See Supplementary Materials for details). Moreover, the dynamical system is symmetric under the change $1 \leftrightarrow 2$, provided we also change the parameter values of the external potentials. A Bell violation occurs whenever
\begin{equation}\label{eq:BellViolation}
|S| > 2.
\end{equation}
Since we avoid parameter values where multistable orbits appear in phase space, the correlations \(C(a,b)\) can be computed from the long-time averages of synchronized trajectories within the manifolds \(S_1\) and \(S_2\). The parameter $|S|$ is then averaged over uniformly distributed initial positions and velocities for the two particles, covering both wells of each double-well potential, thus preventing any bias in the initial conditions. Violations are identified whenever the resulting value of $\langle |S| \rangle$ exceeds the classical bound of 2.

\section{Static Bell violations}

As demonstrated by Papatryfonos {\it et al.}~\cite{PhysRevFluids.9.084001},
classical bipartite systems can violate the CHSH-Bell inequality when the measurement settings are fixed at the start of the experiment and remain unchanged throughout. As discussed in Sec.~\ref{ent:sync}, these entangled states admit a classical interpretation in terms of synchronization between the two droplets, mediated by the wave field. {Likewise}, intermittent transitions between manifolds $S_1$ and $S_2$ produce average correlations or anticorrelations, depending on the well parameters, potentially violating the CHSH bound.

To identify the regions where Bell violations occur in our system, we performed extensive numerical simulations across the parameter space defined by the dimensionless wave amplitude $R$ and the dimensionless memory parameter $\tau$. The resulting correlation functions, $C(a,a)$, $C(a,a^{*})$, and $C(a^{*},a^{*})$, together with the corresponding Bell parameter $|S|$, are shown in Fig.~\ref{fig: 1 in phase}. Although strong correlations and synchronization become increasingly prominent in the high-memory regime [Fig.~\ref{fig: 1 in phase}(a--c)], violations of the Bell inequality are concentrated in the region of moderate memory ($\tau \lesssim 2$) and moderate wave forcing ($R \lesssim 8$), as shown in Fig.~\ref{fig: 1 in phase}(d).

The origin of these violations lies in the coexistence of two distinct dynamical regimes, each governed by a different invariant manifold. Although each of the three correlations may, in principle, be positive or negative depending on the underlying dynamics, violations of the Bell inequality are favored when settings $(a, a)$ and $(a,a^{*})$ are dominated by trajectories evolving near the correlated synchronization manifold $S_1$, while setting $(a^{*},a^{*})$ is dominated by trajectories evolving near the anticorrelated synchronization manifold $S_2$. Therefore, Bell inequality violations occur mainly where this combination of synchronization regimes is realized, corresponding to the blue region in Fig.~3(d). In particular, this region occupies a broad and continuous portion of the parameter space $(R,\tau)$, demonstrating that Bell violations are robust rather than restricted to isolated parameter values. Furthermore, this region of parameter space includes regimes with coexisting attractors, intermittent chaotic dynamics, and fully periodic motion on both synchronization manifolds. The strongest violations occur in the periodic regime, where the trajectories remain confined to the correlated and anticorrelated synchronization manifolds. 

Representative trajectories giving rise to Bell inequality violations are shown in Fig.~\ref{fig: traj 3}. The dynamics exhibits three distinct levels of complexity. First, each droplet undergoes chaotic oscillations within a given potential well. Second, each droplet intermittently switches between the two wells, producing chaotic hopping in the time series shown in Fig.~\ref{fig: traj 3}(a). These two forms of chaos were confirmed by positive Lyapunov exponents and broadband power spectra (see Appendix~\ref{sec:A4}) and are consistent with the chaotic dynamics previously reported for a single walking droplet in a double-well potential~\citep{VALANI2024115253}.
A third level of {complexity} emerges in the $(x_1,x_2)$ phase space projection. As shown in Fig.~\ref{fig: traj 3}(b), the trajectories switch irregularly between the correlated and anticorrelated synchronization manifolds, $S_1$ and $S_2$, both of which are unstable. Since the {system} spends slightly more time near $S_1$ than $S_2$, the resulting correlation is weakly positive. By contrast, in the setting $(a,a^{*})$ (see Figs.~\ref{fig: traj 3}(e,f)), the dynamics exhibit a much stronger preference for $S_1$, resulting in a substantially larger positive correlation.

The situation is qualitatively different in the setting $(a^{*},a^{*})$. Here, the {phase space trajectory} collapses onto the stable anticorrelated synchronization manifold $S_2$, producing perfectly periodic motion (see Figs.~\ref{fig: traj 3}(c,d)). The corresponding phase portrait reduces to a straight line of negative slope, reflecting perfect anticorrelation and yielding $C(a^{*},a^{*})=-1$. The Bell violation therefore arises from the different stability properties of the synchronization manifolds {arising with} the three measurement settings.

Finally, we have computed the CHSH-Bell parameter $S$, which quantifies the degree of Bell violation in this static configuration, and is shown in Fig.~\ref{fig: violation}. The parameter remains positive {for all $n$} and converges robustly to its asymptotic value, $\langle S \rangle = 3.15$. The standard error of the mean was estimated from a sample of initial conditions $n=1000$, which proved sufficient to ensure convergence. As expected, the standard error of the mean $\sigma_{\langle S \rangle}=\sigma/\sqrt{n}$ decreases with increasing sample size $n$, reaching $\sigma_{\langle S \rangle} = 0.0015$. The resulting value therefore establishes a statistically significant violation that exceeds the classical bound by approximately twenty standard deviations $\sigma$. 

\section{Dynamic Bell violations}\label{sec:dynamic_bell}

Beyond static Bell tests, the reduced Lorenz-like model allows the implementation of a fully \emph{dynamic} Bell test protocol {along the lines suggested by Papatryfonos {\it et al.} \cite{PhysRevFluids.9.084001}}. Specifically, we demonstrate that correlations generated by droplet synchronization can violate the CHSH-Bell inequality even when interactions between the two droplets are turned off for arbitrarily long intervals before the measurement settings prescription. The imposition of the measurement settings forces the droplets into one of their two possible states: the system may then be said to undergo `dynamical collapse' \cite{almeida2026gravitationally}. Under these conditions, the requirement of \emph{no-signalling} between the two subsystems is preserved. Although there is no communication between the two subsystems after isolation, the wave field at the time of isolation stores the system's history, and this {wave-mediated} memory is sufficient to preserve the synchronized states and subsequent Bell violations. 

\subsection{Symmetric double wells}

\begin{figure}
    \centering
\includegraphics[width=0.75\columnwidth]{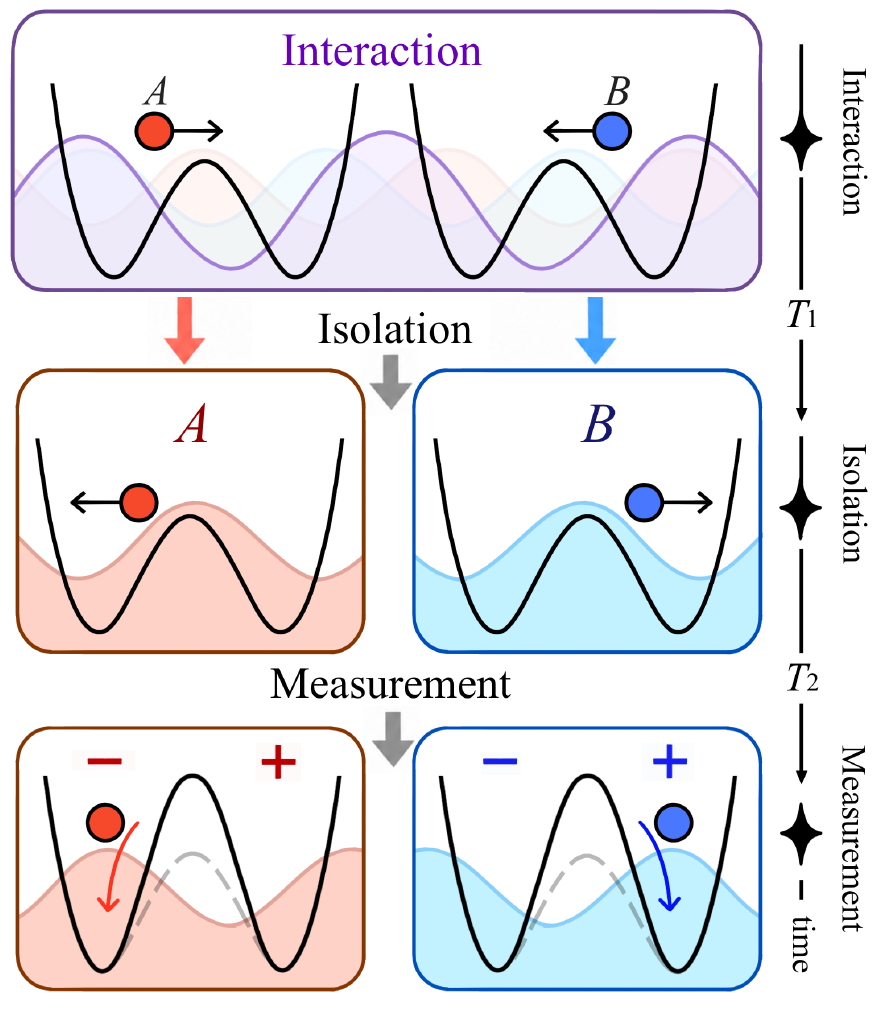}
    \caption{Dynamic Bell-test protocol. The protocol consists of three stages: (i) an interaction stage ($0<T<T_1$), during which the droplets synchronize through wave-mediated interactions; (ii) an isolation stage, initiated at $T_1$ by abruptly switching off the coupling ($\zeta=0$), {during} which each droplet evolves independently; and (iii) a measurement stage, beginning at $T_1+T_2$, where abrupt changes in the double-well potentials define the measurement settings and steer each droplet into one of the two wells, yielding the measurement outcomes.
}
    \label{fig: dynamic protocol}
\end{figure}

We begin by investigating {the possibility of dynamic} violations in the symmetric double-well system introduced in the previous section. For this purpose, we implement a simple measurement protocol 
\cite{guckenheimer2013nonlinear} in which the two subsystems are dynamically decoupled before measurement.
In the parlance of quantum foundations, this decoupling ensures that the ``no-signaling condition'' is satisfied, and that there be no ``conspiratorial correlations'' resulting from communication between the measuring devices.
The proposed protocol consists of three successive stages, as illustrated in Fig.~\ref{fig: dynamic protocol}:

\begin{enumerate}

    \item {The walking droplets become coupled. They} start at a random location within their own double-well potential and evolve for a interaction time $T_1$ with interactions enabled and measurement settings $a$ and $a^*$. This allows the two subsystems to converge to a synchronized state, which may be periodic or chaotic, depending on the well heights.
    
    \item At time $T_1$, the wave-mediated interaction between the droplets is instantaneously switched off (setting $\zeta=0)$. Subsequently, the two subsystems evolve independently for a time $T_2$. 
    
    \item At time $T_1+T_2$, the measurement settings are changed to $b$ and $b^*$ by instantaneously modifying the geometry of the double-well potentials. In particular, we increase the height of the double well, thereby producing a bifurcation in the dynamical system, which results in a fast transient relaxation of each droplet toward one of the two potential minima. This final stage defines the measurement outcomes used to compute the Bell correlations. The duration $T_2$ thus controls the temporal separation between the last interaction and the measurement event.
\end{enumerate}

\begin{figure*}
   \centering
   \includegraphics[width=0.9\columnwidth]{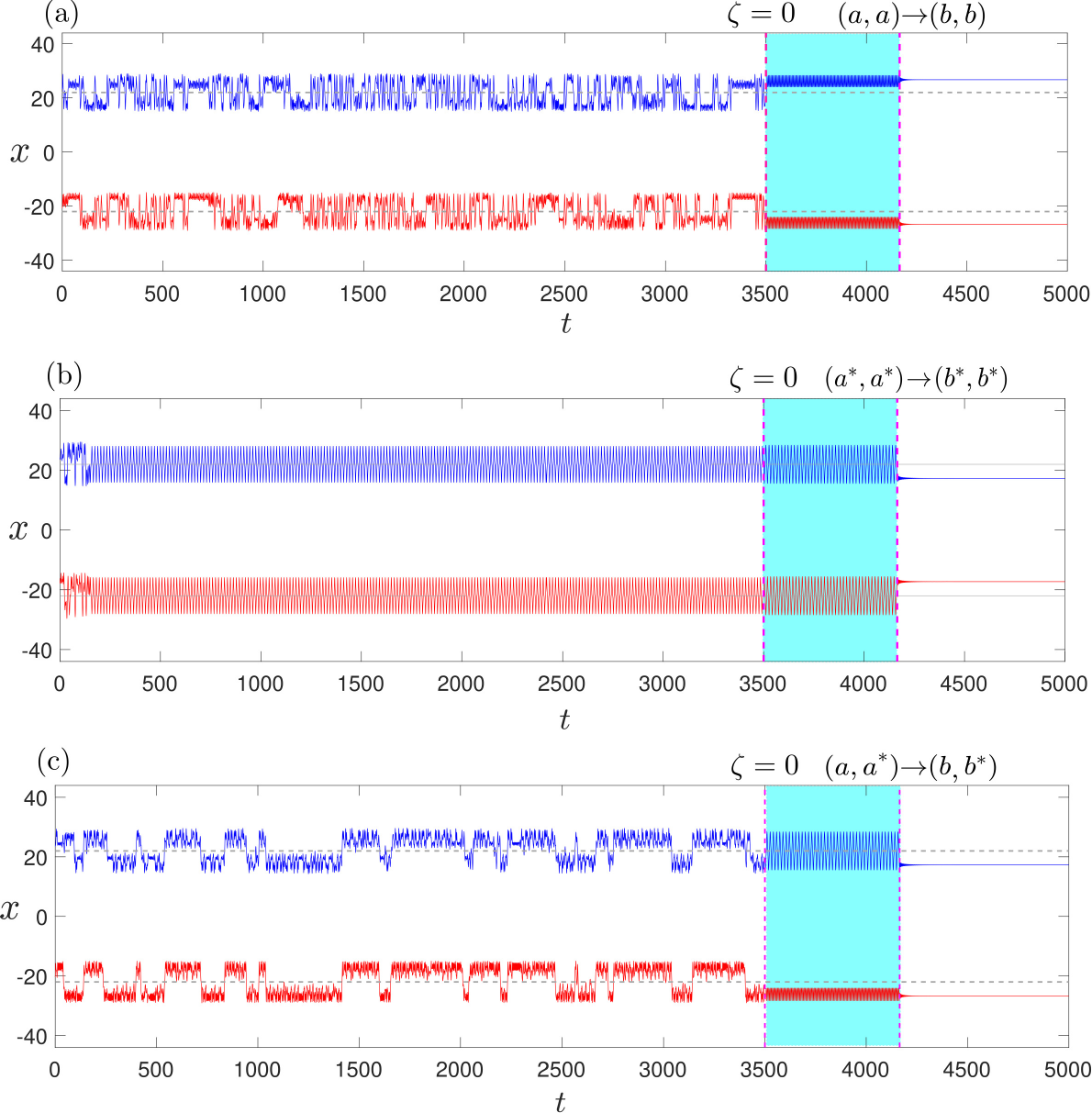}
   \caption{Realization of the dynamical Bell-test protocol yielding a Bell parameter $|S|=2$. Panels (a)--(c) show the time evolution of the droplet positions for the measurement settings $(a, a)$, $(a^{*}, a^{*})$, and $(a, a^{*})$, respectively. During the {coupling} stage ($t<T_1$), the droplets interact through their wave fields and synchronize. At $T_1=3500$ (first dashed magenta line), the interaction is switched off ($\zeta=0$), after which the droplets evolve independently on their self-sustained limit cycles (cyan region). At the measurement time $T_1+T_2=4165$ (second dashed magenta line), the measurement settings are changed from $a \rightarrow b$ and $a^{*} \rightarrow b^{*}$, by increasing the double-well barrier heights. The resulting bifurcation drives each droplet into one of the two wells, thereby defining the measurement outcomes. Parameters are $R=5.29$ and $\tau=0.78$, with $h_1=h_2=6$ for $(a,a)$, $h_1^*=h_2^*=3.5$ for $(a^{*},a^{*})$, $h_1=h_2=20$ for $(b,b)$, and $h_1^*=h_2^*=19$ for $(b^{*},b^{*})$. See also Supplemental Video S3.}
   \label{fig: S4}
\end{figure*}

\begin{figure}
    \centering
    \includegraphics[width=0.75\columnwidth]{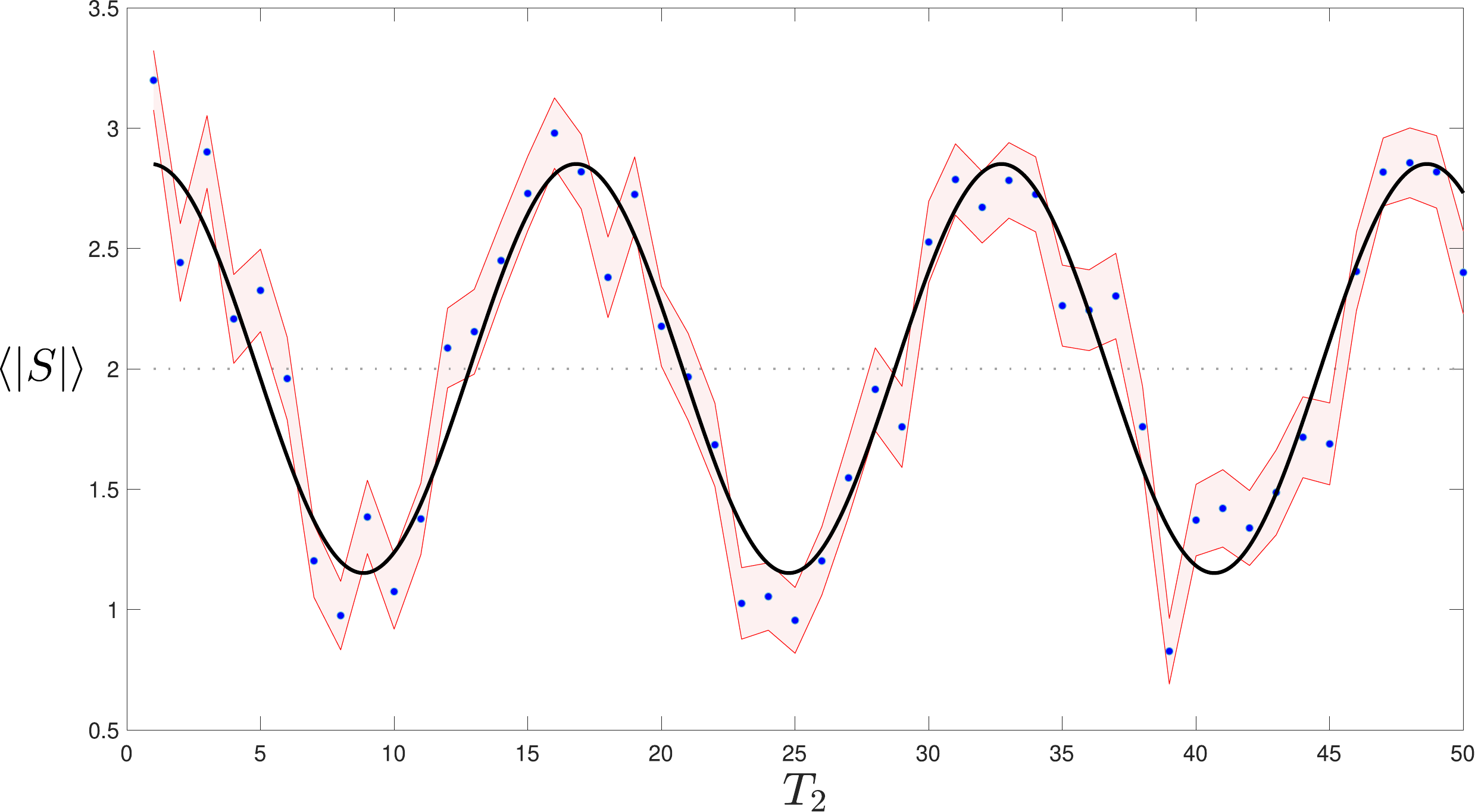}
    \caption{Ensemble-averaged Bell parameter $\langle |S| \rangle$ as a function of the delay time $T_2$, computed from $100$ uniformly distributed, statistically independent initial conditions. The Bell parameter oscillates periodically as the delay time is varied, reflecting the deterministic evolution of the synchronized ensemble during the isolation stage. The red curves denote one standard error of the mean, while the black curve is the least-squares fit, $S(T_2)=S_0+\Delta S\sin(\omega T_2+\phi),$ with $S_0=2.002$, $\Delta S=0.84$, $\phi=1.15$, and period $T=2\pi/\omega=16.2$, corresponding to the characteristic period of the ensemble oscillation between the correlated and anticorrelated synchronization manifolds. The parameters are $R=5.29$ and $\tau=0.78$. The measurement settings are $(a,a)$ with $h_1=h_2=6$ and $(a^{*},a^{*})$ with $h_1^*=h_2^*=3.5$, while the measurement settings are $(b,b)$ with $h_1=h_2=20$ and $(b^{*},b^{*})$ with $h_1^*=h_2^*=19$.}
    \label{fig: dynamic 2}
\end{figure}

Using this protocol, we observe strong violations of the CHSH-Bell inequality for a range of parameters $(R,\tau)$.  
A representative realization of the droplet trajectories in the dynamic Bell test protocol is shown in Fig.~\ref{fig: S4}. During the interaction stage ($t<T_1$), the droplets synchronize via interaction through their common wave field, as arose in the static tests. At $t=T_1$, the interaction is turned off ($\zeta=0$). For $(a, a)$, the chaotic intermittent synchronized state becomes restricted to one of the two wells, whereas for the setting $(a^{*}, a^{*})$ each droplet continues to oscillate autonomously on the anticorrelated limit cycle that encompasses both wells. Finally, at $t=T_1+T_2$, the measurement settings change from $(a, a)$, $(a^{*}, a^{*})$, and $(a, a^{*})$ to $(b, b)$, $(b^{*}, b^{*})$, and $(b, b^{*})$, respectively, by increasing the barrier heights of the double-well potentials. This bifurcation rapidly drives each droplet into one of the two wells, producing a definite measurement outcome.

Applying this protocol to an ensemble of uniformly distributed, statistically independent initial conditions reveals that the CHSH-Bell parameter $S$ is strongly dependent on the delay time $T_2$ between the end of the interaction and the measurement. As shown in Fig.~\ref{fig: dynamic 2}, the ensemble-averaged Bell parameter $\langle |S| \rangle$ exhibits a pronounced sustained oscillation as a function of $T_2$, periodically exceeding the classical CHSH bound of $2$. For the parameter values considered here, the maximum value approaches $\max_{T_2}\langle |S(T_2)| \rangle \approx 2.85$, showing that Bell inequality violations persist even after the wave-mediated interaction between the droplets is turned off. Importantly, these violations persist for arbitrarily long times after the interaction has been shut off. 

One may ask whether a measurement process (corresponding to an objective ``collapse'' mechanism) now needs to be imposed, or whether Bell violations could be detected simply by continuously monitoring the degree of synchronization between the two subsystems, as we did with the static test. In the present model, continuous monitoring reveals strong time-dependent correlations. However, their time average typically does not exceed the CHSH-Bell bound. The introduction of a measurement process — implemented here by rapidly increasing the barrier height of the double-well potentials — selects specific outcomes from the synchronized dynamics, allowing for the realization of a complete protocol with discrete measurement results at a fixed time $T_2$, as is customary in Bell tests \cite{larsson2014loopholes}. From this perspective, the measurement process is not required to generate correlations, but is required to sample them in a manner consistent with the Bell tests. Our system is thus different from its quantum counterpart, in which the delay time $T_2$ is irrelevant.

To further probe the origin of the oscillations evident in Fig.~\ref{fig: dynamic 2}, we note that each realization of the protocol yields only one of the three possible values of the Bell parameter, namely $S\in\{0,2,4\}$, depending on the well the droplets occupy following the measurement-induced bifurcation. Consequently, Bell violations do not necessarily arise at the level of individual trajectories but emerge only after averaging over an ensemble of statistically independent initial conditions. During the interaction stage, depending on measurement settings, synchronization concentrates the ensemble onto a restricted region of phase space with a bias toward one of the synchronization manifolds~\footnote{The probability density of dissipative dynamical systems described by first-order ODEs evolves according to the Liouville equation towards their statistical attracting sets. Physically relevant attracting sets possess a distinguished invariant probability measure---the Sinai--Ruelle--Bowen (physical) measure--- which describes the asymptotic statistics of typical trajectories \cite{eckmann1985ergodic}.}. 

When the interaction is turned off, this ensemble evolves deterministically {according to} the uncoupled dynamics and undergoes a periodic redistribution between the correlated and anticorrelated synchronization manifolds, $S_1$ and $S_2$~(see Supplemental Video S4). As illustrated in Fig.~\ref{fig: S4}(c), this occurs in the mixed setting $(a, a^{*})$ because one droplet evolves along a limit cycle spanning both wells, whereas the other remains confined to a limit cycle within a single well. Consequently, the ensemble periodically changes between the neighborhoods of $S_1$ and $S_2$. Since the measurement process samples the system at a time $T_2$, the relative {prevalence} of these two manifolds determines the probabilities of obtaining the outcomes $S=0$, $2$, and $4$. Therefore, the oscillatory exchange of probability between $S_1$ and $S_2$ gives rise to the periodic variation about 2 of the ensemble-averaged Bell parameter $\langle |S (T_2)| \rangle$ shown in Fig.~\ref{fig: dynamic 2}. Increasing the size of the ensembles ten-fold yields very similar parameter values for the least-squares fitting of the CHSH-Bell parameter, with considerably smaller error bars.
\begin{figure*}
    \centering
    \includegraphics[width=1.0\columnwidth]{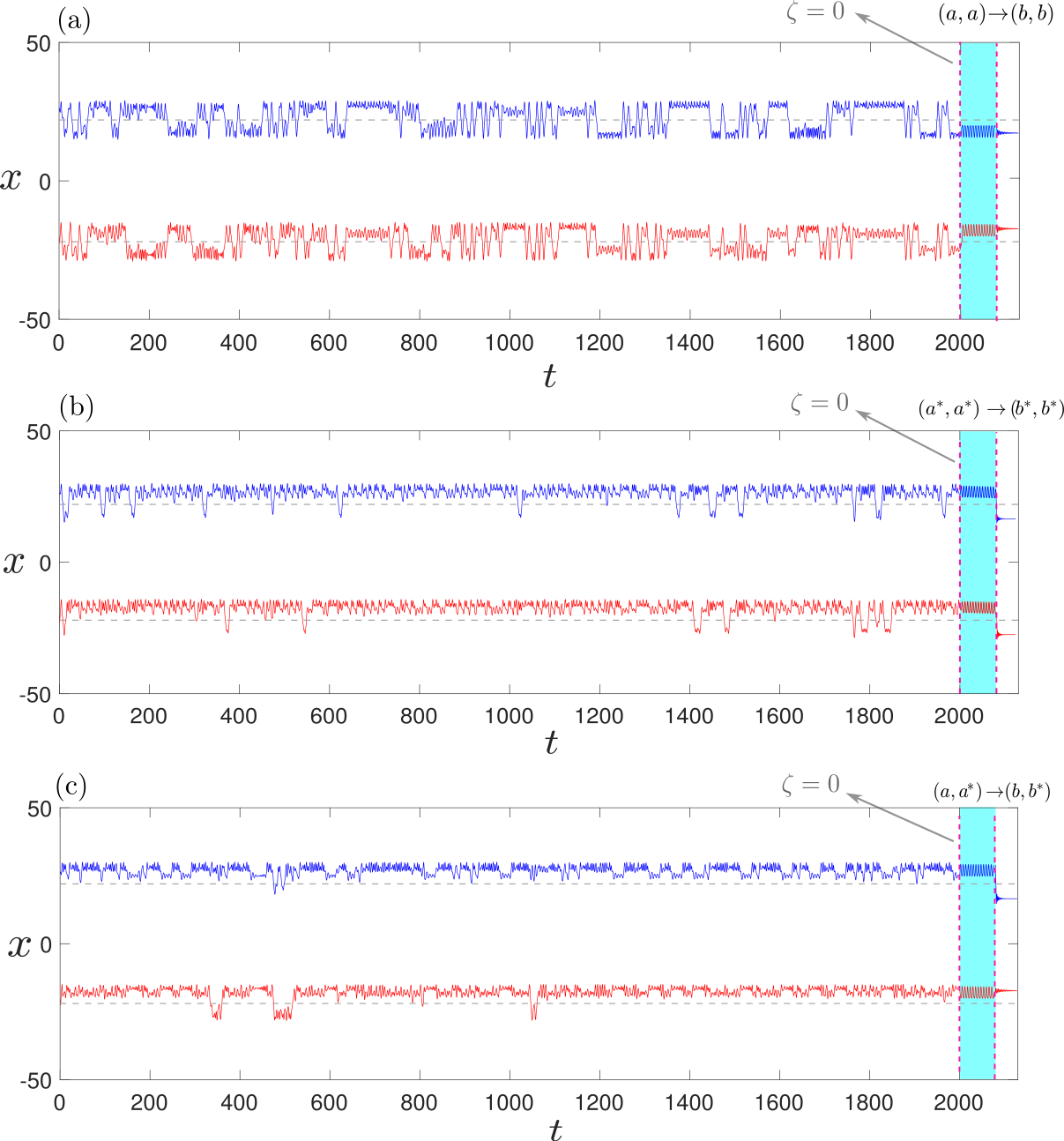}
    \caption{Example of dynamic Bell simulations that give $|S|=4$. Other parameters were fixed to: $R=5.29$ and $\tau=0.78$. Here, setting $(a,a)$ corresponds to $h_1=h_2=6.0$ and $\epsilon_1=\epsilon_2=0$, whereas setting $(a^{*},a^{*})$ corresponds to $h_1^*=h_2^*=3.5$ and $\epsilon_1^*=\epsilon_2^*=1.1$. After {measurement}, the setting $(b,b)$ corresponds to $h_1=h_2=20$ and $\epsilon_1=\epsilon_2=0$, while the setting $(b^{*},b^{*})$ corresponds to $h_1^*=h_2^*=19$ and $\epsilon_1^*=\epsilon_2^*=-6.2$. The first dashed magenta line corresponds to the time of decoupling $T_1=2000$, while the second dashed cyan line corresponds to the time of measurement, $T_1+T_2=2100$. See also Supplemental Video S5.} 
    \label{fig: S4 asymmetry}
\end{figure*}

\subsection{Asymmetric double wells}\label{sec:dynamic_bell_asy}

Although we observe Bell violations at certain {delay times $T_2$, the ensemble average $\langle |S| \rangle$ varies periodically with $T_2$, and} its time-averaged value remains $\langle |S| \rangle_{T_2} = 2$ {(see Fig.~\ref{fig: dynamic 2})}. Therefore, Bell violations are not independent of delay time $T_2$. To overcome this limitation, we now break the symmetry of the double-well potentials experienced by the droplets, following previous studies on Bell violations in walking-droplet systems \cite{PhysRevFluids.9.084001}. This asymmetry reduces the number of synchronization manifolds while biasing both the switching dynamics between the two wells and the measurement process, thereby enabling persistent Bell violations over time. As previously, we employ our measurement protocol to ensure that the no-signaling condition is satisfied.
We modify the double-well potentials confining the droplets as follows
\begin{align}
V_1(x)&=\frac{1}{4}A_1(x+\delta)^4 - \frac{1}{2}B_1(x+\delta)^2 - \epsilon_1 (x+\delta),\\
V_2(x)&=\frac{1}{4}A_2(x-\delta)^4 - \frac{1}{2}B_2(x-\delta)^2 - \epsilon_2 (x-\delta).
\end{align}
The additional linear term introduces a constant tilt to the symmetric double-well potential, rendering it asymmetric, with the degree of asymmetry controlled by the parameters $\epsilon_i$. We ensure that the tilt $\epsilon_i$ never exceeds the critical value $\epsilon_{i,c}=\pm \sqrt{4 B_i^3/27 A_i}$, thus maintaining the confining nature of both wells. 
As shown analytically in the Appendix~\ref{sec:A5}, the imposed asymmetry can induce the bias in the measurement process required to obtain robust Bell violations $\langle |S(T_2)| \rangle>2$ at all $T_2$. We then repeat the simulations using the same protocol described in Sec.~\ref{sec:dynamic_bell}, but with the asymmetric potential.

In Figs.~\ref{fig: S4 asymmetry}(b--c), the chaotic intermittent motion is now biased so that the particles are preferentially confined to one of the wells while still switching to the other at apparently random times. As in the study by Papatryfonos {\it et al.}~\cite{PhysRevFluids.9.084001}, the static Bell parameter $S$ is now significantly reduced because the externally induced asymmetry of the double-well potential leaves only a single synchronization manifold. Consequently, in static Bell tests with asymmetric settings, an imbalance in the transition rates between the two wells is the key mechanism for producing Bell violations~\cite{PhysRevFluids.9.084001}.

The mechanism responsible for the dynamic Bell violations differs from that in the static tests,
since the system neither enters an anticorrelated self-sustained oscillatory regime during synchronization nor does it do so after decoupling. As shown in Fig.~\ref{fig: S4 asymmetry}(a), during the first synchronization stage ($t \leq T_1$), the setting $(a, a)$ remains symmetric, without bias ($\epsilon_1=\epsilon_2=0.0$), and exhibits a moderate correlation, with $C(a, a)$ fluctuating around $0.5$. By contrast, as shown in Fig.~\ref{fig: S4 asymmetry}(b), the setting $(a^{*}, a^{*})$ is biased toward the right well by choosing $\epsilon_1^*=\epsilon_2^*=1.1$. The setting $(a, a^{*})$, with $\epsilon_1=0.0$ and $\epsilon_2^*=1.1$, is also biased, due to the wave-mediated coupling between the two walking droplets (see Fig.~\ref{fig: S4 asymmetry}(c)).

We can exploit these asymmetric intermittent states to achieve sustained Bell violations during {any} interval $T_2$ by inducing a {biased measurement process}. To this end, when the measurement settings are changed from $a \rightarrow b$ we keep the setting without tilt ($\epsilon_i=\epsilon_i=0.0$), while for the change $a^{*} \rightarrow b^{*}$, the tilt is changed from $\epsilon^{*}_i=1.1$ to $\epsilon^{*}_i=-6.2$, where $i=\{1,2\}$. As shown in Fig.~\ref{fig: S4 asymmetry}, this induces a positive correlation for the $(a^{*}, a^{*})$ setting, since both droplets collapse in their leftmost wells. {Conversely}, for the $(a,a^{*})$ setting, the droplets remain biased toward their respective rightmost wells throughout the wave-mediated coupling. Consequently, the measurement process induces an anticorrelation in this case because the first droplet, whose potential is unbiased, now collapses into the rightmost well~(see Supplemental Video S6). The two correlation functions $C(a,a^{*})$ and $C(a^{*},a^{*})$ already yield an average CHSH-Bell parameter of $S=3$. Finally, because the droplets in the $(a, a)$ setting remain partially synchronized, switching chaotically between the two synchronization manifolds, most realizations of this dynamical test give $S=2$ or $S=4$, with only a comparatively small fraction delivering $S=0$. Such events occur when random transitions take place in the $(a,a^{*})$ configuration. 

\begin{figure}
    \centering
    \includegraphics[width=0.8\columnwidth]{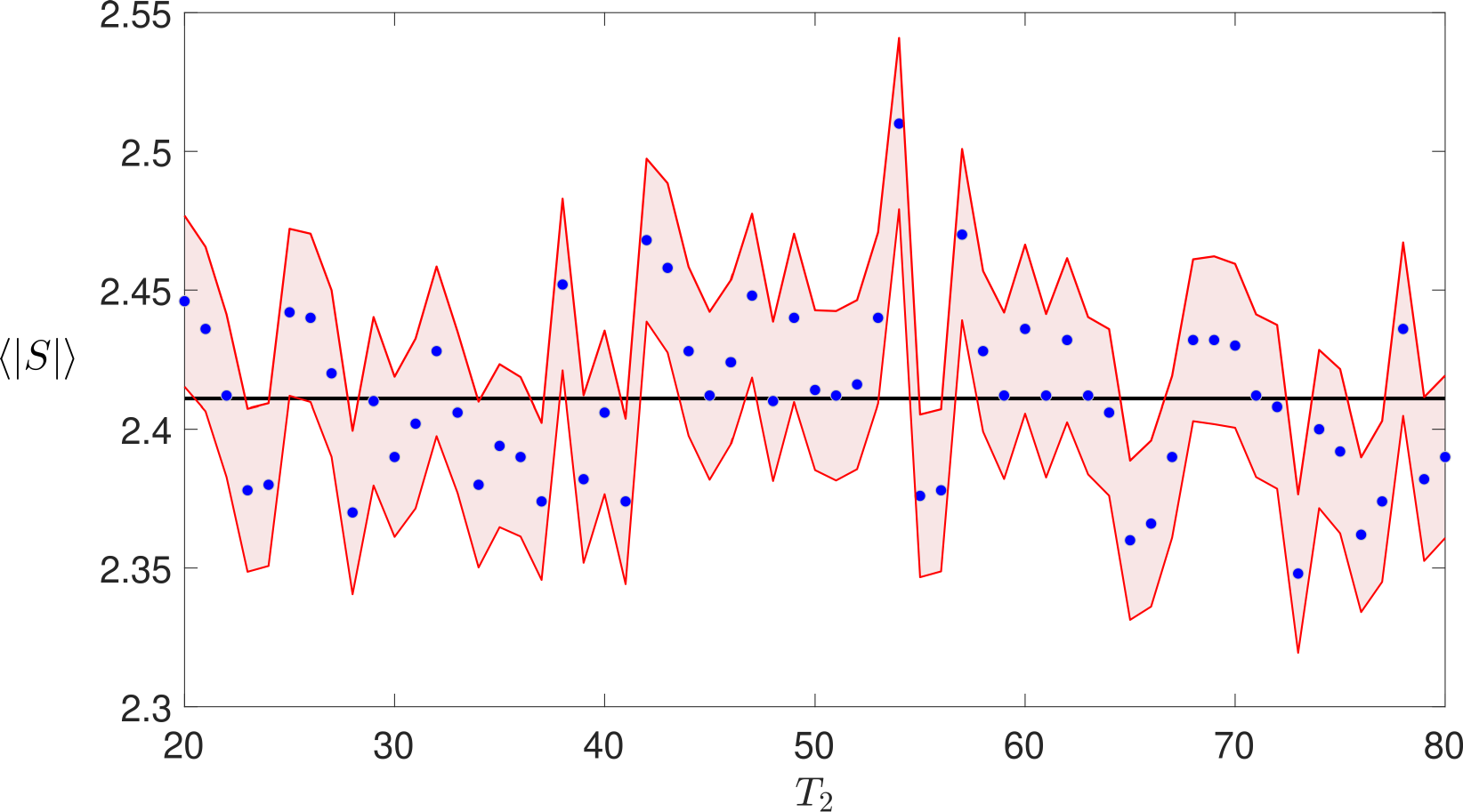}
    \caption{Variations of average Bell parameter $\langle |S| \rangle $ with delay time $20 \le T_2 \le 80$ calculated from an ensemble of $1000$ initial conditions. The red curves represent the standard error of the mean, while the black line corresponds to the average {value over the time interval shown}. Here, the setting $(a,a)$ corresponds to $h_1=h_2=6.0$ and $\epsilon_1=\epsilon_2=0.0$, whereas the setting $(a^{*},a^{*})$ corresponds to $h_1^*=h_2^*=3.5$ and $\epsilon_1^*=\epsilon_2^*=1.1$. After the collapse, the setting $(b,b)$ corresponds to $h_1=h_2=20$ and $\epsilon_1=\epsilon_2=0.0$, while and the setting $(b^{*},b^{*})$ corresponds to $h_1^*=h_2^*=19$ and $\epsilon_1^*=\epsilon_2^*=-6.2$.}
    \label{fig: dynamic 2 asymmtry}
\end{figure}

As shown in Fig.~\ref{fig: dynamic 2 asymmtry}, the resulting CHSH-Bell violations are sustained over time and fluctuate with the delay time $T_2$, with a temporal average of $\langle |S|\rangle_{T_2} \approx 2.42$. We may thus conclude that wave-mediated synchronization of intermittent dynamics together with the measurement process can produce sustained Bell violations at any delay time $T_2$. 

\section{Discussion and Conclusions}

We have demonstrated that strong Bell violations can emerge {from a} deterministic classical pilot-wave system through nonlinear synchronization mechanisms. Using a reduced Lorenz-like model derived from the integro-differential dynamics of walking droplets, we have shown that coupled wave-mediated dynamics naturally induce correlated and anticorrelated synchronization manifolds, intermittent transitions between which yield CHSH-type Bell violations over broad regions of parameter space. Importantly, these violations may be rationalized in terms of chaotic pilot-wave dynamics without invoking stochastic quantum jumps, non-local interactions, or intrinsically probabilistic hidden variables.

The central dynamical mechanism underlying the observed correlations is the synchronization of periodic switching and chaotic intermittency. In the strongly nonlinear regime, trajectories evolve in the vicinity of invariant synchronization manifolds associated with correlated and anticorrelated droplet motion, intermittently transitioning between them due to manifold instability and the underlying chaotic phase-space structure. The resulting \emph{multiscale intermittency} produces long-time statistical correlations whose degree of entanglement depends on the relative stability and {\it prevalence} of the synchronization manifolds. From this perspective, entanglement and Bell violations emerge dynamically from the geometry of the underlying strange attractor structure.

The reduced Lorenz-like formulation provides a particularly transparent dynamical interpretation of the {Bell violations. In the formulation of Papatryfonos {\it et al.}~\cite{PhysRevFluids.9.084001}, the drop dynamics were described in terms of partial differential equations that define an infinite-dimensional dynamical system. Conversely}, the finite-dimensional representation resulting from our
choice of wave form $W(x) = \cos{x}$
reveals how invariant manifolds, bifurcations, and synchronization stability organize the correlated dynamics. In particular, the existence of two symmetry-based synchronization manifolds associated with the reflection group provides a natural explanation for the coexistence of correlated and anticorrelated states. The present results demonstrate that systems with higher-order symmetry groups can exhibit a larger number of synchronization subspaces, enabling the engineering of bipartite systems and the entanglement of their dynamics in a variety of ways ~\footnote{Specifically, quantum systems and hydrodynamic quantum analogs in higher-dimensional configuration spaces usually possess continuous Lie groups of symmetries (\emph{e. g.}, the spin of entangled fermions described by the Dirac equation transforms according to representations of the SU(2) Lie group. In contrast, Eq.~\eqref{eq:dimensional_particle} in three dimensions has O(3) symmetry \cite{kay2025classical}), which can provide a non-denumerable set of synchronization manifolds.}.

The dynamical Bell protocol introduced here is important in that it extends previous walking-droplet Bell tests \cite{PhysRevFluids.9.084001}. Earlier studies considered static Bell tests in which the interacting subsystems remained continuously coupled through the wave field. The present protocol explicitly turns off droplet–droplet interactions before the measurement stage, then implements a measurement procedure by rapidly modifying the confining potentials. The persistence of Bell-type correlations following decoupling demonstrates that synchronized states may survive long after direct interaction has ceased, due to the nature of the coupling function and the self-sustained oscillatory dynamics of individual droplets \cite{Nachbin2018,jenkins2013self}. In this respect, the system exhibits a form of dynamical memory analogous to persistent phase synchronization in coupled nonlinear oscillators \cite{correlationnachbin}. In short, our system demonstrates that memory (specifically, time-delay coupling) can account for entanglement \cite{Bush2010}.

An important conceptual aspect of the present work concerns the role of measurement independence. As in the prior hydrodynamic Bell tests \cite{PhysRevFluids.9.084001}, the pilot-wave system does not satisfy the assumption of measurement independence used to derive the CHSH-Bell inequality \cite{Vervoort2018,hossenfelder2020rethinking}. The hidden variables of the system include the extended wave field, whose form depends on the geometry of the confining potentials {(or measurement settings)} through their influence on the particle's trajectory. This measurement dependence persists after the two particles are uncoupled and is also responsible for the oscillations of the correlation parameter $S$. Consequently, the probability distribution over hidden variables cannot, in general, be factorized independently of the detector configuration \cite{quantum7030029}, and therefore $p(\lambda) \neq p(\lambda|a,b)$ in Eq.~\eqref{eq:28}. 

Our results suggest several promising directions for investigating Bell violations in classical pilot-wave systems. We emphasize that the pilot-wave model adopted here, based on a sinusoidal wave kernel, was deliberately chosen to maximize wave-mediated coupling. In walking-droplet systems, however, the wave kernel is more realistically described by a spatially decaying Bessel function \cite{Damiano2016,Couchman2019}. Furthermore, these waves are generated by successive droplet impacts and propagate at finite speed, rather than instantaneously throughout the domain. An important open question is whether Bell violations can also arise in a fully dynamical, local pilot-wave model, such as that developed by Milewski {\it et al.}~\cite{Milewski2015} and subsequently adapted by Papatryfonos {\it et al.}~\cite{PhysRevFluids.9.084001} for static Bell tests. Our study suggests that Bell violations in their system might be most judiciously sought through consideration of the dynamics of their system's synchronization manifolds.

The broader significance of the present work lies in the connection it establishes between Bell-type correlations, synchronization theory, and classical pilot-wave dynamics. Synchronization phenomena are ubiquitous across nonlinear science, appearing in systems ranging from neural dynamics and chemical oscillators to coupled lasers and fluid instabilities \cite{pikovsky1985universal, Strogatz2000, Cvitanovic2020Chaos}. The fact that synchronization, together with the measurement-dependent pilot-wave form \cite{hossenfelder2020rethinking}, can generate Bell-type statistical correlations suggests that certain structures commonly regarded as uniquely quantum may instead reflect more general organizational principles of complex multiscale nonlinear dynamical systems involving deterministic fields and chaotic dynamics \cite{l3xp-yrrv}. In this sense, the hydrodynamic quantum analogs \cite{BushOza2020} continue to provide a valuable conceptual framework to explore the fundamental aspects of quantum {theory}.

\section*{Acknowledgements}
R.V. acknowledges the support of the Leverhulme Trust [Grant No. LIP-2020-014]. JB acknowledges valuable discussions with Andr\'{e} Nachbin, and the financial support of the ONR through grant N000014-24-1-2232.

\clearpage
\appendix

\begin{center}
\textbf{Appendix: Synchronization induces Bell violations in a model of walking droplets}
\end{center}

\vspace{1cm}

\noindent\textbf{Summary.}
In this Appendix, we provide the mathematical derivations underlying the synchronization manifolds, synchronization conditions, and dynamical Bell oscillations discussed in the main text.

\section{Reduced formulation of the coupled Lorenz-like system}

\subsection{Reduction from the 12D to the 8D system}

In the main text, coupled walking-droplet dynamics is formulated
as a 12D dynamical system, with each droplet described by its
position $x_i$, velocity $X_i$, and two pairs of memory variables,
$(Y_{ii},Z_{ii})$ and $(Y_{ij},Z_{ij})$, representing the self-wave and cross-wave contributions, respectively. For the synchronization
analysis presented here, it is convenient to introduce an equivalent
8D formulation by combining both contributions for each droplet.

Specifically, for the coupled system ($\zeta=1$), we define
\begin{equation}
    Y_1 = Y_{11}+Y_{12},
    \qquad
    Z_1 = Z_{11}+Z_{12},
    \label{eq:Y1Z1_reduction}
\end{equation}
and
\begin{equation}
    Y_2 = Y_{22}+Y_{21},
    \qquad
    Z_2 = Z_{22}+Z_{21}.
    \label{eq:Y2Z2_reduction}
\end{equation}
Thus, $Y_i$ represents the total wave-mediated memory force acting on droplet $i$, while $Z_i$ represents the corresponding total wave
amplitude evaluated at the droplet position. Adding the evolution equations for $Y_{11}$ and $Y_{12}$, and likewise
those for $Z_{11}$ and $Z_{12}$, allows us to transform the twelve-dimensional Lorenz-like system of the main text to the eight-dimensional reduced dynamical system
\begin{align}
    \dot{x}_1 &= X_1, \nonumber\\
    \dot{X}_1 &= -X_1+Y_1+F_1(x_1), \nonumber\\
    \dot{Y}_1 &= -\frac{Y_1}{\tau}
                 +X_1Z_1
                 +R\sin(x_1-x_2), \nonumber\\
    \dot{Z}_1 &= 2R-\frac{Z_1}{\tau}
                 -X_1Y_1
                 +R(\cos(x_1-x_2)-1),
    \label{eq:reduced_system_1}
\end{align}
and
\begin{align}
    \dot{x}_2 &= X_2, \nonumber\\
    \dot{X}_2 &= -X_2+Y_2+F_2(x_2), \nonumber\\
    \dot{Y}_2 &= -\frac{Y_2}{\tau}
                 +X_2Z_2
                 +R\sin(x_2-x_1), \nonumber\\
    \dot{Z}_2 &= 2R-\frac{Z_2}{\tau}
                 -X_2Y_2
                 +R(\cos(x_2-x_1)-1).
    \label{eq:reduced_system_2}
\end{align}

This reduction introduces no additional approximation since, for
$\zeta=1$, the equations for the summed memory variables close exactly. Therefore, the 8D formulation contains the same coupled
dynamics as the 12D formulation, but provides a more compact representation for the synchronization analysis below, following previous works \cite{VALANI2024115253}. In particular, each droplet may now be represented by the four-dimensional state vector
\begin{equation}
    \mathbf{x}_i=(x_i,X_i,Y_i,Z_i),
\end{equation}
which is the notation employed in the following sections.

\subsection{Coupled dynamical-system formulation}

We begin by considering a generic coupled dynamical system of the form
\begin{align}
\dot{x}_1 &= F(x_1) + H(x_2-x_1), \\
\dot{x}_2 &= F(x_2) + H(x_1-x_2),
\end{align}
where the vector field $F$ defines the intrinsic dynamics of each subsystem and $H$ is a nonlinear coupling function. Complete synchronization corresponds to the invariant manifold $ x_1 = x_2$. Substituting this condition into the equations of motion yields
\begin{align}
\dot{x}_1 &= F(x_1)+H(0),\\
\dot{x}_2 &= F(x_1)+H(0).
\end{align}

Thus, whenever the dynamical systems are identical and couple symmetrically, the synchronization manifold exists as an invariant subspace of the full phase space. In addition, if $H(0)=0$, this subspace is maintained after the decoupling of the particles. More generally, generalized synchronization corresponds to a nonlinear relation
\begin{equation}
x_2 = G(x_1),
\end{equation}
where $G$ is a differentiable transformation. Differentiating with respect to time gives
\begin{equation}
\dot{x}_2 = DG(x_1)\dot{x}_1.
\end{equation}
Substituting the equations of motion yields
\begin{align}
DG(x_1)\Big[F(x_1)+H(G(x_1)-x_1)\Big]
=&\ F(G(x_1))
\nonumber \\
&+ H(x_1-G(x_1)).
\end{align}
Therefore, generalized synchronization manifolds satisfy the following requirements
\begin{equation}
\begin{split}
F(G(x_1))
- DG(x_1)F(x_1)
={}& DG(x_1)H(G(x_1)-x_1)
\\
&- H(x_1-G(x_1)).
\end{split}
\label{eq:8}
\end{equation}
In the case that the transformation $G$ is linear, we can split this condition into the two independent conditions $F=G^{-1}FG$ and $ H=GHP$, where $P$ reflects the vector $x$ in phase space, and the second identity is evaluated in the synchronization manifold.

For the Lorenz-like walking-droplet model discussed in the main text, the relevant symmetry transformation is
\begin{equation}
G(x,X,Y,Z)=(-x,-X,-Y,Z),
\label{eq:9}
\end{equation}
which follows from the reflection symmetry of the reduced dynamical system. The vector field of the reduced model is
\begin{align}
F_x &= X, \\
 F_X &= Y-X-Ax^3+Bx,\\
 F_Y &= -\frac{Y}{\tau}+XZ,\\
 F_Z &= 2R-XY-\frac{Z}{\tau},
\end{align}
while the coupling field is
\begin{align}
H_x &=0,\\
H_X &=0,\\
H_Y &= R\sin(x),\\
H_Z &= R(\cos(x)-1).
\end{align}
Using the symmetry transformation above, one readily verifies that Eq.~\eqref{eq:8} holds when the dynamical system $F$ and the coupling $H$ are considered, together with the symmetry transformation of Eq.~\eqref{eq:9}, which proves the existence of the generalized synchronization manifold.

\section{Synchronization manifolds and symmetry groups} \label{sec:A2}

We now establish a more general theorem relating synchronization manifolds to the symmetry structure of coupled dynamical systems.

\subsection{Theorem on synchronization manifolds}

The following general theorem can be used to prove the existence of two synchronization manifolds in our Lorenz-like system.

\textbf{Theorem 1.} Let $\Phi$ be a dynamical system defined on a phase space $Y=X\times X$ of dimension $2N$, consisting of two identical dynamical systems of dimension $N$ coupled symmetrically through a nonlinear interaction function $H$. Suppose that the uncoupled dynamics $F$ is invariant under the action of a symmetry group $G$, and the coupling satisfies
\begin{equation}
H = gHP,
\end{equation}
with $P=-I$ and $g$ a faithful linear representation of the symmetry group. Then, there exist $|G|$ invariant synchronization manifolds given by
\begin{equation}
x_2=g(x_1), \qquad \forall g\in G.
\end{equation}

\subsection{Proof}
The coupled system can be written as
\begin{align}
\dot{x}_1 &= F(x_1)+H(x_2-x_1),\\
\dot{x}_2 &= F(x_2)+H(x_1-x_2).
\end{align}
Evaluating the dynamics on the synchronization manifold $x_2=g(x_1)$ gives
\begin{align}
\dot{x}_1 &= F(x_1)+H(g(x_1)-x_1),\\
Dg\,\dot{x}_1 &= F(g(x_1))+H(x_1-g(x_1)).
\end{align}

Multiplying the first equation by $Dg$ yields
\begin{equation}
Dg\dot{x}_1 = DgF(x_1)+DgH(g(x_1)-x_1).
\end{equation}

Comparing both expressions gives the consistency condition
\begin{equation}
F(g(x_1))+H(x_1-g(x_1))
= DgF(x_1)+DgH(g(x_1)-x_1).
\end{equation}

If the vector field is invariant under the symmetry group ($F=g^{-1}Fg,$), then the existence condition reduces to
\begin{equation}
H(x_1-g(x_1))=gH(-(x_1-g(x_1))),
\end{equation}
and $H=gHP$. Thus, the number of synchronization manifolds is directly determined by the order of the symmetry group of the dynamical system.

\subsection{Stability of synchronization manifolds}

Define a perturbation away from the synchronization manifold by
\begin{equation}
z=x_2-G(x_1).
\end{equation}
Linearization around the synchronization manifold, together with the conditions for $F$ and $H$ derived in the previous section, yields
\begin{equation}
\dot{z}=s(t)z,
\end{equation}
where, in the case that $G$ is linear, the stability matrix is
\begin{equation}
s(t)=DF(G(x_1))-D\tilde{H}(x_1),
\end{equation}
where $D\tilde{H}(x_1)=DH(x_1-G(x_1))+G DH(G(x_1)-x_1)$. Synchronization is stable whenever the maximum transverse Lyapunov exponent of the stability matrix is negative. 

\section{Analytical synchronization conditions for the Lorenz-like droplet model} \label{sec:A3}

We now derive stability synchronization conditions for the reduced Lorenz-like walking-droplet model.
\subsection{Complete synchronization}
Consider the coupled system
\begin{align}
\dot{x}_1 &= F(x_1)+H(x_2-x_1),\\
\dot{x}_2 &= F(x_2)+H(x_1-x_2),
\end{align}
with $H(0)=0$. Define the synchronization error
\begin{equation}
z=x_2-x_1.
\end{equation}
The evolution equation for the synchronization error is
\begin{align}
\dot{z}
&= F(x_1+z)-F(x_1)+H(-z)-H(z).
\end{align}
Linearization around $z=0$ yields
\begin{equation}
\dot{z}=[DF(x_1)-2DH(0)]z+\mathcal{O}(\norm{z}^2).
\end{equation}
Introducing the transformed variable
\begin{equation}
w=e^{2DH(0)t}z,
\end{equation}
one obtains
\begin{equation}
\dot{w}=DF(x_1(t))w.
\end{equation}
Therefore,
\begin{equation}
\norm{w}\sim Ce^{\lambda_{\max} t},
\end{equation}
where $\lambda_{\max}$ is the maximum Lyapunov exponent of the uncoupled system. For example, in the standard case where the coupling is diffusive, synchronization is guaranteed whenever
\begin{equation}
DH(0)>\lambda_{\max}/2.
\end{equation}

However, since our coupling field $H$ is non-trivial, the criterion for stability is given by the maximum Lyapunov exponent of the matrix
\begin{equation}
s(t)=
\begin{pmatrix}
0 &1&0&0\\
-3Ax_1^2+B & -1 & 1 &0\\
-2R&Z_1&-1/\tau&X_1\\
0&-Y_1&-X_1&-1/\tau
\end{pmatrix}.
\end{equation}

\subsection{Generalized synchronization}

For generalized synchronization $x_2=G(x_1)$, with
\begin{equation}
G=
\begin{pmatrix}
-1 &0&0&0\\
0&-1&0&0\\
0&0&-1&0\\
0&0&0&1
\end{pmatrix},
\end{equation}
define the synchronization error $z=x_2-G(x_1)$. Linearization around the generalized synchronization manifold yields
\begin{equation}
\dot{z}=s(t)z,
\end{equation}
where the Jacobian matrix is
\begin{equation}
s(t)=
\begin{pmatrix}
0 &1&0&0\\
-3Ax_1^2+B & -1 & 1 &0\\
0&Z_1&-1/\tau&-X_1\\
0&Y_1&X_1&-1/\tau
\end{pmatrix}.
\end{equation}

Its trace is
\begin{equation}
\mathrm{Tr}(s)=-(1+2/\tau),
\end{equation}


{which shows that the generalized synchronization dynamics is dissipative, with the phase-space contraction rate decreasing as the memory parameter $\tau$ increases. This condition alone does not determine the stability of the synchronization manifold, which instead requires the maximum transverse Lyapunov exponent associated with $\mathbf{s}(t)$ to be negative.}

\section{Unpredictability of the synchronized chaotic intermittent motion} \label{sec:A4}

The Lyapunov exponent spectrum provides a quantitative characterization of the asymptotic stability of trajectories in a dynamical system. For an $n$-dimensional flow, the spectrum consists of $n$ Lyapunov exponents,
\[
\lambda_1 \geq \lambda_2 \geq \cdots \geq \lambda_n,
\]
which measure the average exponential rates of growth or decay of infinitesimal perturbations along mutually orthogonal directions in phase space. In particular, the maximum Lyapunov exponent is defined as
\begin{equation}
\lambda_{1}
=
\lim_{t\rightarrow\infty}
\frac{1}{t}
\ln\left(
\frac{\|\delta\mathbf{x}(t)\|}
{\|\delta\mathbf{x}(0)\|}
\right),
\end{equation}
where $\delta\mathbf{x}(t)$ is an infinitesimal perturbation that evolves according to the variational equations associated with the dynamical system. Thus, $\lambda_{1}$ represents the average exponential rate of divergence (or convergence) of the nearby trajectories. A positive value of $\lambda_{1}$ is the defining signature of deterministic chaos, indicating sensitive dependence on initial conditions and is therefore a hallmark of chaotic dynamics, while the sum of all exponents gives the average phase-space volume contraction rate in dissipative systems. 
\begin{figure*}
    \centering
    \includegraphics[width=1.0\columnwidth]{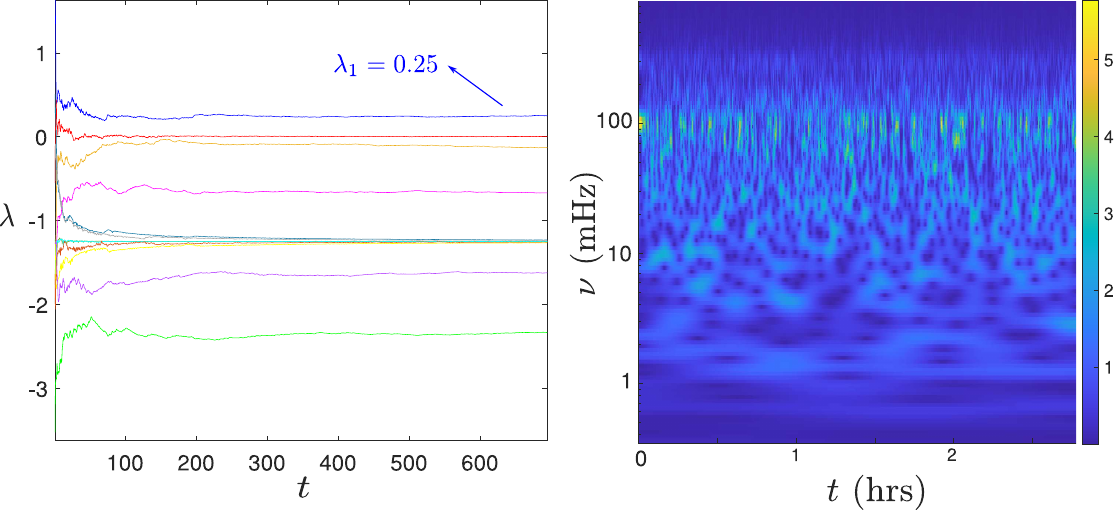}
    \caption{Unpredictability of chaotic intermittency for the trajectories of the static Bell violations at $R=5.44$ and $\tau=0.8$, and setting $(a, a)$ corresponding to $h_1=h_2=6$. (a) The Lyapunov spectrum shows one positive Lyapunov exponent $\lambda_1=0.25$, confirming chaotic dynamics. (b) The wavelet transform of the time series shows a complex continuous spectrum that fluctuates in time. These two hallmarks of chaos show that the evolution is unpredictable in practice, despite the correlations between the two droplets.}
    \label{fig: lyap}
\end{figure*}
The Lyapunov spectrum was computed using the algorithm introduced by Wolf \emph{et al.}, which simultaneously integrates the nonlinear system with its variational equations \cite{Wolf1985}. At regular time intervals, the tangent vectors are orthonormalized using the Gram-Schmidt procedure to prevent numerical overflow while accumulating the logarithmic stretching factors. The time-averaged growth rates of these orthogonal perturbation vectors converge to the full Lyapunov exponent spectrum as the integration time becomes sufficiently long. This method has become the standard numerical approach for evaluating Lyapunov exponents in continuous dynamical systems. The full Lyapunov spectra can be seen in Fig.~\ref{fig: lyap}(a).

As a second measure of deterministic chaos, we have computed the wavelet transform of the chaotic time series. The time-dependent frequency content of the signal was analyzed using the continuous wavelet transform (CWT), which can be regarded as a localized generalization of the Fourier transform. By correlating the signal with wavelets of different scales and positions, the CWT produces a time-frequency map that tracks the evolution of the dominant oscillatory modes. The wavelet scalograms presented here were generated using MATLAB's \texttt{cwt} routine, and are depicted in Fig.~\ref{fig: lyap}(b).

\section{Oscillations of the Bell parameter after isolation} \label{sec:A5}

In this section, we analyze the origin of the oscillatory Bell correlations discussed in the main text. Consider two periodic signals with random phases, representing idealized oscillations of the particle around the homoclinic points that separate the two wells of the double-well. Then, the values of our dichotomous variable can be written as
\begin{equation}
x_1(t)=\mathrm{sgn}[\cos(\omega t+\phi_1)],
\end{equation}
and
\begin{equation}
x_2(t)=\mathrm{sgn}[\cos(\omega' t+\phi_2)].
\end{equation}

The correlation function is
\begin{equation}
\begin{split}
\langle{x_1(t)x_2(t)}\rangle
={}&
\int_0^{2\pi}
\mathrm{sgn}\big[\cos(\omega t+\phi_1)\big]
\mathrm{sgn}\big[\cos(\omega' t+\phi_2)\big] p(\phi_1,\phi_2)
\,d\phi_1 d\phi_2,
\end{split}
\end{equation}
where $p(\phi_1,\phi_2)$ is the probability density of hidden initial phases for the droplets, which generally depends on the synchronization state. 

\subsection{Perfect synchronization}

For complete synchronization $C_{a a}=\langle x_1(t)x_2(t)\rangle$, the probability density resulting from the synchronization process can be written as
\begin{equation}
p_{aa}(\phi_1,\phi_2)
=\delta(\phi_1-\phi_2)p(\phi_1).
\end{equation}
yielding the correlation integral
\begin{equation}
\begin{split}
C_{a a}
={}&
\int_0^{2\pi}
\mathrm{sgn}\big[\cos(\omega t+\phi_1)\big]
\mathrm{sgn}\big[\cos(\omega' t+\phi_2)\big] p_{aa}(\phi_1,\phi_2)
\,d\phi_1 d\phi_2 .
\end{split}
\end{equation}
\vspace{10px}
If we now consider $\omega=\omega'$, we get the result
\begin{equation}
C_{aa}=1.
\end{equation}
Note how the measurement dependence is introduced here through the synchronization process, by setting $p_{aa}(\phi_1,\phi_2)$. Similarly, for the anticorrelated synchronization manifold, we get
\begin{equation}
p_{a^{*}a^{*}}(\phi_1,\phi_2)
=\delta(\phi_1-\phi_2-\pi)p(\phi_1).
\end{equation}
and therefore we get the result
\begin{equation}
C_{a^*a^*}=-1.
\end{equation}

\subsection{Independent phases}

If the phases are statistically independent, we have for the crossed correlation 
\begin{equation}
p_{a a^{*}}(\phi_1,\phi_2)=p_1(\phi_1)p_2(\phi_2),
\end{equation}
and then, if we further assume that the probabilities are uniformly distributed, with $p_i(\phi_i)=1/2\pi$, we obtain the result
\begin{equation}
C_{aa^*}=0,
\end{equation}
and consequently $S=2$. 
Thus, within this reduced phase-oscillator description, obtaining a
nonzero crossed correlation, and hence exceeding $S=2$ in the present
construction, requires correlations between the phases, nonuniform
phase distributions, or asymmetry of the underlying oscillations.\\

\subsection{Phase-locked oscillations}

Assuming correlated phases of the form
\begin{equation}
p_{aa^*}(\phi_1,\phi_2)
=\delta(\phi_2-\phi_1-\phi_0)p(\phi_1),
\end{equation}
one obtains
\begin{align}
&    \Delta S(t) = 2\int_0^{2\pi}
\mathrm{sgn}
\left[
\cos(\omega t+\phi_1)
\cos(\omega' t+\phi_1+\phi_0)
\right]
p(\phi_1)d\phi_1.
\end{align}

The Bell parameter then becomes
\begin{equation}
S(t)=2+\Delta S(t),
\end{equation}
which oscillates periodically or quasiperiodically depending on the commensurability of the frequencies $\omega$ and $\omega'$. This result implies that signals that produce the same subtended area below and above zero provide oscillations that give $\langle \Delta S \rangle_t = 0$. Introducing asymmetric oscillations, for example through the function $x_1(t)=\mathrm{sgn}[\epsilon_1+\cos(\omega t+\phi_1)]$
and $x_2(t)=\mathrm{sgn}[\epsilon_2+\cos(\omega' t+\phi_2)]$, and appropriately collapsing the system, we can obtain time-averaged violations. If the values of $\epsilon_1$ and $\epsilon_2$ are not too big to destroy the existence of two wells for each double-well, one finds a nonzero average contribution $\langle{\Delta S(t) \rangle}_t$, which depends on the values of $\epsilon_1$ and $\epsilon_2$. Hence, sustained and time-averaged Bell violations can be readily obtained by introducing asymmetries in \emph{both} effective double-wells.

\bibliography{Sync_bell}

\end{document}